\documentclass[preprint,showpacs,preprintnumbers,amsmath,amssymb, superscriptaddress,longbibliography,nofootinbib]{revtex4-1}
\usepackage{graphicx} % Required for inserting images
\usepackage{xcolor}
\usepackage{amsmath}
 \usepackage{hyperref}
\usepackage[top = 2cm, bottom = 2cm, right = 2cm, left =2cm]{geometry}
\usepackage{enumitem}

\usepackage{tikz}
\usetikzlibrary{decorations.pathmorphing,calc}
\begin{document}

\title{Geometric Saturation of the Scale Function in Black-Bounce Spacetimes: Spherical, Planar, and Hyperbolic Transverse Sections}

\author{Milko Estrada }
\email{milko.estrada@gmail.com}
\affiliation{Departamento de Física, Facultad de Ciencias, Universidad de Tarapacá, Casilla 7-D, Arica, Chile}

\date{\today}

\begin{abstract}
We present a new construction for black-bounce spacetimes based on a deformation of the scale function that acts as a geometric saturation rather than introducing a prescribed constant minimal length, as in conventional black-bounce models. This function encodes the gravitational information of the underlying vacuum counterpart, keeping the geometric quantities of the deformed metric finite in the short-distance region where tidal forces and curvature invariants of the vacuum geometry diverge. The deformation saturates at short scales, suppressing curvature divergences without requiring a potentially unstable de Sitter core. The resulting regular geometries with spherical, planar, and hyperbolic transverse sections describe regular black holes (RBHs), extremal RBHs, and traversable wormholes. A key result is that the deformed geometry remains finite at $l=0$ in all regimes, while, depending on the parameter space, the bounce, identified with the minimum of $R(l)$, may remain in the short-distance region or shift to a larger finite value of $l$. Thus, the deformation regularizes the central region and, in some regimes, also modifies the global spacetime profile. Spherical and planar RBHs satisfy the standard energy conditions near the bounce, showing that a geometric bounce does not necessarily require exotic matter sources in this region. In the hyperbolic case, the energy conditions depend more strongly on the mass parameter, being satisfied near the bounce for positive-mass RBHs but violated for negative-mass and extremal configurations. The hyperbolic sector also admits regular negative-mass black holes. For wormhole geometries, the WEC and NEC are violated near the throat, as expected.

\end{abstract}

\maketitle

\section{Introduction}

In recent years, observations such as the detection of gravitational waves from the merger of rotating black holes \cite{LIGOScientific:2016aoc}, together with the images of the centers of the galaxies M87 \cite{EventHorizonTelescope:2019dse} and Sagittarius A* \cite{EventHorizonTelescope:2019ggy}, have shown remarkable agreement with the predictions of General Relativity (GR), providing strong observational evidence for the existence of black holes (BHs). These observations have opened the possibility of directly probing the strong-gravity regime and testing different geometric proposals beyond the standard classical scenarios. In this context, a particularly relevant line of research aims to understand to what extent effective modifications of the space-time geometry at short scales could alter the internal structure of black holes without modifying their expected asymptotic behavior.

Despite their phenomenological success, the usual BH solutions in GR contain physical singularities where curvature invariants diverge and the classical description of spacetime ceases to be valid. From a physical viewpoint, these singularities are associated with divergent tidal forces, implying that any extended object reaching them would inevitably be destroyed. In this connection, it is well known that tidal-force effects can be characterized through the Kretschmann invariant \cite{Bena:2020iyw}. In a vacuum spacetime with constant-curvature transverse sections, this quantity behaves as $K_v \sim r^{-6}$, implying that tidal forces diverge at short distances near the origin. In this way, the curvature invariant can be interpreted as an indicator that the geometry of a vacuum BH has entered the high-curvature regime at short length scales. Within this framework, regular black holes (RBHs) have been widely studied as effective models designed to avoid the appearance of physical singularities at short or Planckian scales \cite{Spallucci:2017aod}. This observation motivates considering effective deformations of the vacuum BH geometry through a deformation function constructed from curvature invariants of the underlying vacuum spacetime. Such a function is assumed to act in a saturating manner, becoming more relevant in the high-curvature regime while remaining almost negligible at large distances. As a consequence, the resulting effective geometry remains regular and its curvature invariants stay finite at short length scales. A well-known example that can be interpreted within this perspective is the Dymnikova model (DyM) \cite{Dymnikova:1992ux}. From this perspective, this BH can also be viewed as an effective deformation of the Schwarzschild vacuum geometry. In this case, the Newtonian potential is effectively screened by a screening function (SF) that depends on the Kretschmann invariant, of the form $\mathrm{SF}=1-\exp(-a/K_v^{\,1/2})$, where $a$ is a constant, yielding the effective metric $g_{tt}=-g_{rr}^{-1}=1-\frac{2M}{r}\,\mathrm{SF}$. As a result, the screening function becomes significant in the strong-field regime at short scales, suppressing the central singularity and yielding a regular effective geometry. It is worth mentioning that the Dym model belongs to the class of RBHs with a de Sitter core. Such cores have been identified as potentially unstable in some studies \cite{Carballo-Rubio:2022pzu}, motivating the search for alternative mechanisms of geometric regularization. It is worth mentioning that, within a very different framework, the matter sources associated with the effective geometry of the DyM model have attracted considerable interest due to their formal analogy with the pair-production factor appearing in the Schwinger effect in quantum electrodynamics \cite{DymnikovaS1996,Ansoldi:2008jw}, with more recent studies along these lines reported in Refs.~\cite{Alencar:2023wyf,Estrada:2023pny,Estrada:2024uuu,Errehymy:2025djk,Errehymy:2026ylh}.  It is worth emphasizing that the aforementioned analogy with the Schwinger effect is beyond the scope of the present work and requires a more detailed investigation.

From a broader perspective, effective deformations of the Schwarzschild geometry have also been formulated independently of specific regular black-hole models. In this direction, Ref. \cite{Konoplya:2022hll} proposed an effective description of black-hole geometries through deformation functions depending on physical quantities associated with the spacetime, such as the proper distance, the Ricci scalar, and the Kretschmann scalar. Since curvature invariants provide natural measures of the local gravitational field, they constitute suitable quantities for characterizing the onset of strong-curvature effects. This idea also appears in renormalization-group approaches to asymptotically safe gravity, where curvature invariants are employed to identify the running energy scale associated with quantum corrections. In particular, Ref. \cite{Borah:2026sja} relates the Kretschmann scalar to the renormalization-group scale through $K_v \propto k^{4}$, while Ref. \cite{Ishibashi:2021kmf} employs an analogous scale identification to construct renormalization-group improved black-hole geometries. In these approaches, the deformation becomes relevant in the high-curvature regime while remaining almost negligible at large distances, recovering the classical solution asymptotically and, for the constructions of Refs. \cite{Borah:2026sja, Ishibashi:2021kmf}, yielding regular effective geometries at short length scales. It is also worth mentioning that, in Ref. \cite{Platania:2019kyx}, renormalization-group improved Schwarzschild black holes are obtained through a scale-dependent Newton's gravitational coupling. The resulting screening function is closely related to that of the Dymnikova model and can therefore also be expressed in terms of the Kretschmann scalar of the underlying vacuum geometry, as discussed above. The above discussion serves only to motivate the use of the Kretschmann scalar to deform vacuum geometries in physically motivated scenarios; the aforementioned frameworks are beyond the scope of the present work. Another example of effective metric deformations can be found in Ref.~\cite{DelPiano:2024gvw}, where the Schwarzschild geometry is described through deformation functions depending on spacetime quantities such as the proper distance, the Ricci scalar, and the Kretschmann scalar. This self-consistent framework parametrizes deformations of the classical geometry while preserving its symmetries and unifies different deformation prescriptions within a common formalism.

On the other hand, within the study of deformations of the Schwarzschild metric leading to the emergence of nonsingular compact objects, the original black-bounce model proposed by Simpson and Visser (SV) in Ref.~\cite{Simpson:2018tsi} has attracted considerable attention due to its ability to smoothly interpolate between RBHs and traversable wormholes. This model is characterized by a spherically symmetric transverse section of the form $R(l)^2d\Omega^2$, where $l\in(-\infty,\infty)$ is the extended radial coordinate, to be defined below, and the scale function is given by $R(l)^2=l^2+b_0^2$, with $b_0$ a constant. Through an embedding procedure, the scale function is identified with the usual Euclidean radial coordinate according to $r\rightarrow R(l)$. As a consequence, the Euclidean radial coordinate acquires a minimum value $b_0$, such that $r\in[b_0,\infty)$. From this perspective, the SV construction can be interpreted as an effective deformation of the Schwarzschild vacuum geometry in which the radial coordinate is replaced by a scale function incorporating a constant minimal length. In this model, the causal structure does not terminate at a physical singularity but instead extends into another future causal region of spacetime. Depending on the parameter space, these geometries may describe one-way or two-way traversable wormholes, as well as RBHs, without necessarily requiring potentially unstable de Sitter cores. Since its introduction, the black-bounce model has been generalized in several directions, and numerous physical aspects have been investigated, including gravitational lensing \cite{Nascimento:2020ime}, gravitational-wave echoes \cite{Yang:2021cvh}, electromagnetic properties of accretion disks \cite{Bambhaniya:2021ugr}, tidal forces \cite{Arora:2023ltv}, and several other recent applications \cite{Nascimento:2025mtr,Ling:2025ncw,Bragado:2025jrg,Santos:2025xbk,Lessa:2024erf,Crispim:2024yjz}.

From a more general perspective, there has also been growing interest in the theoretical study of static pseudo-spherically symmetric spacetimes with hyperbolic or planar transverse sections, associated with both wormholes \cite{Lobo:2009du,Avalos:2025hfw} and black holes \cite{Calza:2025mrt,Nagasaki:2019icm,Ren:2019lgw,Guo:2020zqm,Sadeghi:2020bon}. In particular, black holes with hyperbolic sections exhibit nontrivial causal properties, since they may possess event horizons even for negative values of the mass parameter \cite{Mann:1997jb,Mann:1997iz}. There also exist examples with nonconstant transverse sections \cite{Yang:2023nnk,Aros:2023tbh,Hull:2021bry}, although this scenario lies beyond the scope of the present work.

Motivated by the considerations discussed above, in the present work we propose an effective construction for black-bounce spacetimes based on a curvature-dependent modification of the scale function of a vacuum black-hole solution. Unlike the original Simpson--Visser (SV) model, where the scale function is controlled by a constant minimal length, we assume that the short-scale modification is instead governed by an effective screened curvature indicator constructed from the Kretschmann invariant of the corresponding vacuum geometry. More specifically, we preserve the geometric structure underlying the SV construction, in which the departure from the vacuum radial geometry is introduced through the scale function $R(l)$. Here, instead of introducing a prescribed constant minimal length, we promote this contribution to a curvature-dependent quantity carrying information about the corresponding vacuum geometry, and consider the identification $R(l)^2=l^2+F_{\rm eff}$. To construct $F_{\rm eff}$, we define $F\sim K_v^{1/2}$ as the curvature indicator of the corresponding vacuum geometry. In this way, $F$ encodes the growth of the vacuum curvature and diverges at short distances when the Kretschmann invariant of the vacuum counterpart becomes unbounded. We then introduce the effective screened curvature indicator through the phenomenological relation $F_{\rm eff}=F\cdot\mathrm{SF}$, with a screening function analogous to that appearing in the Dymnikova model discussed above, $\mathrm{SF}=\bar{A}(1-\Gamma)$, where $\Gamma\sim\exp(-F_c/F)$. Here, $\bar{A}$ is a dimensional constant with $[\bar{A}]=\ell^4$, where $\ell$ denotes length, so that $[F_{\rm eff}]=\ell^2$, while $F_c$, with $[F_c]=\ell^{-2}$, characterizes the onset of the saturation regime. In the Dymnikova construction, the screening acts directly on the Schwarzschild potential through $2M/r\rightarrow(2M/r)(1-\Gamma)$. Here, instead, we investigate the geometric effects of incorporating the screened curvature indicator into the scale function. As will be seen below, since $R(l)$ is identified with the Euclidean radial coordinate and also enters the metric potential $A(R(l))$, this prescription modifies both the scale function itself and the corresponding gravitational potential. The role of the screening function is precisely to convert the divergent behavior of the vacuum curvature indicator at short distances into a finite geometric correction: when $F\gg F_c$, one has $F_{\rm eff}\rightarrow \bar{A} F_c$, whereas in the weak-curvature regime $F\ll F_c$, $F_{\rm eff}\rightarrow0$. Consequently, the deformation becomes relevant where the corresponding vacuum curvature would diverge, while remaining almost negligible asymptotically. As will be shown below, when incorporated into the scale function, this finite saturated correction gives rise to a regular black-bounce geometry.The exponential form adopted here, analogous to that of the Dymnikova model, should be regarded as a phenomenological realization of these limiting properties rather than as a unique prescription derived from an underlying fundamental theory. As will be seen below, the resulting deformation preserves the asymptotic behavior of the vacuum solution while generating a finite scale function at short distances. Depending on the parameter space, the construction gives rise to regular black holes, extremal regular black holes, and traversable wormholes. We additionally investigate how this curvature-dependent black-bounce construction is affected when the transverse section has constant spherical, planar, or hyperbolic curvature.

\section{Physical and geometrical analysis}

\subsection{Transverse section}

We start by describing the line element of a two-dimensional transverse section
\begin{equation}
ds_T^2 = \frac{d\rho^2}{1 - k\rho^2} + \rho^2 d\phi^2,
\end{equation}
where $k = 1,-1, 0$. As we will see below, $k$ represents the curvature
of the surface: $k = 1$ to a spherical geometry (positive curvature), $k = -1$ corresponds to a hyperbolic geometry (negative curvature), and $k = 0$ to a plane geometry.

We perform the following change of variable:
\begin{equation}
\rho =
\begin{cases}
\sin\theta & \text{for } k=+1 \\
\sinh h  & \text{for } k=-1 \\
p  & \text{for } k=0 
\end{cases}
\end{equation}
which leads to the following form for the transverse section:

\begin{equation} \label{ElementoTransversal}
ds_{T-k}^2 =
\begin{cases}
d\theta^2 + \sin^2\theta\, d\phi^2 & \text{for } k=+1 \\
dh^2 + \sinh^2 h\, d\phi^2 & \text{for } k=-1 \\
dp^2 + p^2\, d\phi^2 & \text{for } k=0 
\end{cases}
\end{equation}

In this way, the transverse section of the last equation represents a sphere $\mathcal{S}^2$, a hyperboloid $\mathcal{H}^2$, or a plane $\mathcal{R}^2$. In the spherical case, the transverse metric is defined with coordinate ranges $0 \le \theta \le \pi$ and $0 \le \phi < 2\pi$. In the hyperbolic case, the transverse metric is defined with coordinate ranges $-\infty < h < \infty$ and $0 \le \phi < 2\pi$. In the planar case, the transverse metric is defined with coordinate ranges $0 \le p < \infty$ and $0 \le \phi < 2\pi$. Thus, the usual two-dimensional transverse spheres are replaced by two-dimensional transverse pseudo-spheres. As mentioned in the introduction, in general, any static spacetime can be written in a form corresponding to a transverse section of constant curvature spherical, planar, or hyperbolic, describing a non-trivial compact object such as a wormhole or a black hole.

\subsection{The Spacetime}

As we will see later, we will study black-bounce–type geometries capable of reproducing black holes, such that the Killing horizon (determined by the condition $g_{00}=0$) and an outer marginally trapped surface (the causal horizon, determined by the condition $g_{11}^{-1}=0$) coincide. This is achieved directly (among other possibilities) in the case where $g_{00} = -(g_{11})^{-1}$. Thus, we begin by analyzing the simplest case, given by the following line element

\begin{equation} \label{Metrica1aVersion}
ds^2 = -A\left ( R(l) \right )\,dt^2 + \frac{\,dl^2}{A\left ( R(l) \right )} + R(l)^2 \cdot ds_{T-k}^2
\end{equation}
where $t$ is a temporal coordinate representing the proper time of a static observer, and $l$ represents the proper distance at fixed $t$ also referred to as the extended--radial coordinate. The function $R(l)$ can be interpreted as a scale function of the transverse section of constant curvature, spherical, hyperbolic, or planar. In agreement with the previous subsection, the coordinates range as follows:

\begin{align}
   \mbox{For $k=1$:} &\,\,-\infty < t < \infty, \quad -\infty < l < \infty, \quad \theta \in [0,\pi], \quad \phi \in [0,2\pi] \label{RangoEsferico} \\
   \mbox{For $k=-1$:} &\,\,-\infty < t < \infty, \quad -\infty < l < \infty, \quad -\infty < h < \infty, \quad \phi \in [0,2\pi] \label{RangoHiperboloide} \\
   \mbox{For $k=0$:} &\,\,-\infty < t < \infty, \quad -\infty < l < \infty, \quad 0 \le p < \infty, \quad \phi \in [0,2\pi] \label{RangoPlanar}
\end{align}

\subsection{The generic scale function}

We perform the embedding of the metric into a three-dimensional Euclidean spacetime. In this way, we assume a spatial hypersurface at constant $t$. For this, for simplicity, we use the following assumptions:
\begin{align}
  \mbox{For $k=1$:} &\,\,\,\,  \theta=\theta_0=\pi/2 \\
  \mbox{For $k=-1$:} &\,\,\,\,  h=h_0=\sinh^{-1} (1) \\
  \mbox{For $k=0$:} &\,\,\,\,  p=p_0=1 
\end{align}
First, we can note that the induced metric of the transverse section, independently of the value and sign of the curvature $k$, depends only on the coordinate $\phi$, such that $ds_{T-k}^2 \to d\phi^2$. Thus, the induced metric becomes (at $t=\text{const}$):
\begin{equation} \label{MetricaInducida}
ds^2 = \frac{dl^2}{A\left (R(l) \right )} + R(l)^2\, d\phi^2,
\end{equation}

Now we embed this geometry into a three-dimensional Euclidean space with cylindrical coordinates $(z,r,\phi)$:
\begin{equation} \label{MetricaEuclidea}
ds^2 = dz^2 + dr^2 + r^2 d\phi^2.
\end{equation}
By comparing Eqs. \eqref{MetricaInducida} and \eqref{MetricaEuclidea}, while keeping the extended radial coordinate constant, we can identify the Euclidean radius as
\begin{equation} \label{RelacionRadial}
r = R(l)
\end{equation}

It is straightforward to notice that, for non-constant $l$, by comparing Eq. \eqref{RelacionRadial}, together with $dR = \frac{dR}{dl}\, dl$, to the Euclidean metric, one can integrate the form of the Euclidean cylindrical coordinate $z(l)$. It is worth mentioning that in the usual black-bounce case \cite{Simpson:2018tsi}, where $r=R(l)=\sqrt{l^2+b_0^2}$ with $b_0$ a constant, and consequently $l=\pm\sqrt{r^2-b_0^2}$, the domain of the Euclidean radius is $r\in[b_0,\infty)$, whereas the domain of the proper coordinate is $l\in]-\infty,\infty[$, in agreement with Eqs.\eqref{RangoEsferico}, \eqref{RangoHiperboloide} and \eqref{RangoPlanar}. 

\section{Curvature-Dependent Black-Bounce Construction}

As discussed in the Introduction, the Kretschmann invariant of the corresponding vacuum solution provides a natural measure of the local gravitational field and the onset of the high-curvature regime. For vacuum geometries with constant-curvature transverse sections, we define the curvature indicator
\begin{equation}\label{F}
F\sim\sqrt{K_v}\sim\frac{\bar{M}}{l^3},
\end{equation}
where $\bar{M}$ is related to the mass parameter of the vacuum solution. In geometrized units, $[\bar{M}]=\ell$, where $\ell$ denotes length, and consequently $[F]=\ell^{-2}$. Thus, $F$ encodes the curvature information of the corresponding vacuum geometry and becomes unbounded at short distances when the vacuum Kretschmann invariant diverges.

Our aim is to map this divergent vacuum curvature behavior into a finite geometric correction in the high-curvature regime, while preserving the vacuum behavior asymptotically. To this end, we consider a screening prescription satisfying the following properties: the correction should become almost negligible in the weak-curvature regime, remain finite when the vacuum curvature indicator becomes unbounded, and provide a smooth interpolation between both regimes. Following the phenomenological curvature-dependent screening functions discussed in the Introduction, a simple realization of these requirements is given by
\begin{equation}
\mathrm{SF}=\bar{A}(1-\Gamma),
\end{equation}
where
\begin{equation}
\Gamma\sim\exp(-F_c/F).
\end{equation}
Here, $F_c$ is a constant characterizing the onset of the saturation regime, with $[F_c]=\ell^{-2}$, while $\bar{A}$ is a dimensional constant with $[\bar{A}]=\ell^4$. We then define the effective screened curvature indicator
\begin{equation}\label{Fefectivo}
F_{\rm eff}
=
F\cdot\mathrm{SF}
=
F\cdot \bar{A}\cdot(1-\Gamma),
\end{equation}
which satisfies $[F_{\rm eff}]=\ell^2$, as required for its subsequent incorporation into the squared scale function. To simplify the subsequent numerical analysis and focus on the physically relevant parameters, hereafter we fix the magnitude of the dimensional constant to unity, $\bar{A}=1\,[\ell^4]$.

The exponential screening function adopted above is not assumed to represent a unique prescription. Rather, it constitutes a phenomenological realization of the required limiting behavior, belonging to the same exponential class of curvature-dependent screening functions discussed in the Introduction. Its relevant property for the present construction is that the divergent behavior encoded by $F$ is converted into a finite saturated quantity at high curvature, while its contribution progressively vanishes in the weak-curvature regime.

Indeed, in the weak-curvature regime, corresponding to large distances, one obtains $F_{\rm eff}\approx \bar{A} \cdot F$, whereas in the high-curvature regime the effective screened curvature indicator approaches the finite value $\bar{A} \cdot F_c$. Defining
\begin{equation}
F_c=\frac{\bar{M}}{a^3},
\end{equation}
where $a$ is a constant with dimensions of length, Eq.~\eqref{Fefectivo} becomes
\begin{equation}\label{FefectivoReemplazado}
F_{\rm eff}
=
\bar{A}\cdot\frac{\bar{M}}{l^3}
\left[
1-\exp\left(-\frac{l^3}{a^3}\right)
\right].
\end{equation}

From this expression, it follows that in the high-curvature regime, corresponding to short length scales $l\ll a$ (or equivalently $F\gg F_c$),
\begin{equation}
F_{\rm eff}
=
\bar{A} F_c
\left[
1-\frac{1}{2}\left(\frac{l}{a}\right)^3
+\mathcal{O}\!\left(\left(\frac{l}{a}\right)^6\right)
\right],
\end{equation}
and therefore
\begin{equation}
F_{\rm eff}
\longrightarrow
\bar{A} F_c
=
\frac{\bar{A}\bar{M}}{a^3}
\qquad
(l\rightarrow0).
\end{equation}
Thus, although the curvature indicator $F$ of the corresponding vacuum geometry becomes unbounded in this regime, the effective screened curvature indicator approaches a finite value.

On the other hand, in the weak-curvature regime, corresponding to large distances $l\gg a$ (or $F\ll F_c$),
\begin{equation}
F_{\rm eff}
\simeq
\bar{A} F(l\rightarrow\infty)
\rightarrow
0.
\end{equation}

Therefore, the curvature-dependent correction becomes almost negligible asymptotically.

We now incorporate this finite screened quantity into the black-bounce geometry. The choice of modifying the scale function is motivated by the geometric structure of the original SV construction. In that model, the departure from the vacuum radial geometry $R(l)^2=l^2$ is introduced precisely through the scale function according to $R(l)^2=l^2+b_0^2$, where $b_0$ is a prescribed constant minimal length. Here, we preserve this geometric prescription but replace the constant contribution $b_0^2$ by the effective screened curvature indicator $F_{\rm eff}$, which carries information about the curvature regime of the corresponding vacuum geometry. We therefore propose
\begin{equation}
\label{FuncionEscala1}
R(l)
=
\sqrt{l^2+F_{\rm eff}}.
\end{equation}
In this sense, the relation between $F_{\rm eff}$ and $R(l)$ should be understood as a curvature-dependent extension of the SV geometric prescription, rather than as a unique relation derived from the Kretschmann invariant. Since $[F_{\rm eff}]=\ell^2$, both contributions entering $R(l)^2$ have the same dimensions.

Consequently, the divergent curvature behavior of the vacuum counterpart is encoded through $F$ and screened into a finite correction to the scale function in the high-curvature regime, while the correction becomes negligible at large distances,
\begin{equation}
\label{FormaGeneralDeR}
R(l)^2=
\begin{cases}
\text{short scales:} &
l^2+\text{finite saturation correction},
\\[1mm]
\text{large scales:} &
l^2+\text{almost negligible correction}
\simeq
l^2.
\end{cases}
\end{equation}
As will be shown below, this saturated behavior gives rise to a regular black-bounce geometry while preserving the asymptotic behavior of the corresponding vacuum solution.

In this work, we investigate the resulting black-bounce geometries. Depending on the parameter values, the spacetime may describe a regular black hole, an extremal regular black hole, or a traversable wormhole. Since the conserved energy associated with a timelike Killing vector is well defined for asymptotically AdS spacetimes, we consider geometries described by
\begin{equation}
\label{FuncionA}
A\big(R(l)\big)
=
k+\frac{R(l)^2}{L^2}
-\frac{2M}{R(l)}.
\end{equation}
where $L$ plays a role analogous to the AdS radius in the asymptotic regime. In the next sections, we analyze the behavior of $R(l)$ for different values of the parameter space and for spherical, hyperbolic, and planar transverse sections. Furthermore, we derive the corresponding energy--momentum tensor and analyze the energy conditions.

\subsection{Parameter Space and Geometrical Black Bounce Structure}

We study the regions of the parameter space in which the black-bounce model leads to a regular black hole (extremal or non-extremal) or to a wormhole. We begin by analyzing the behavior of the scale function $R(l)$, which is identified with the Euclidean radial coordinate $r$, with the bounce corresponding to its minimum value. As discussed above, in the high-curvature regime the vacuum curvature indicator $F$ becomes unbounded, while the effective screened curvature indicator approaches the finite value $F_{\rm eff}\rightarrow \bar{A} F_c$. Consequently, the scale function remains finite at the origin, $R(0)=\sqrt{\bar{A} F_c}$, in all the regimes considered.

\begin{figure}[h]   
    \centering 
    \includegraphics[scale=.85]{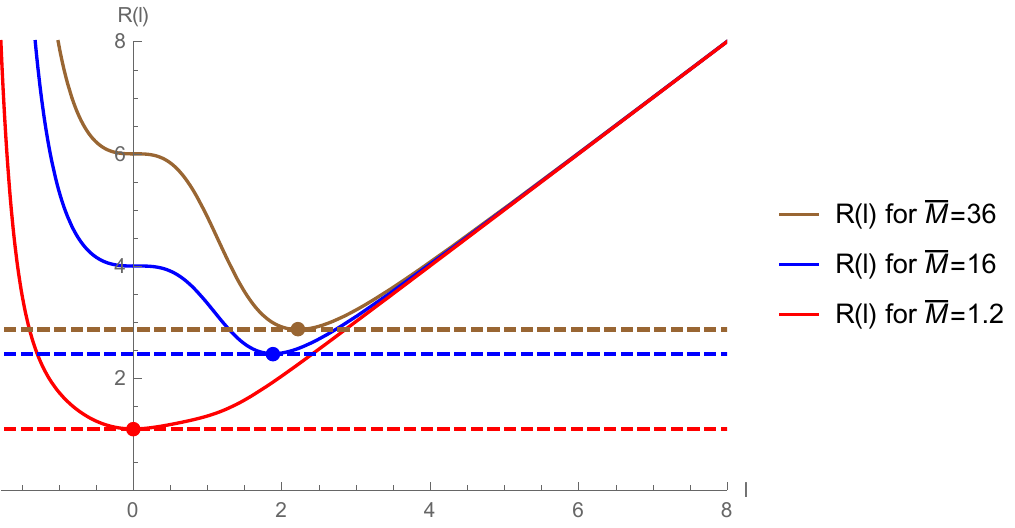} 
    \caption{$R(l)$ for $a=1$ and $\bar{M}=36,16,1.2$, shown in brown, blue, and red, respectively. The minimum points $(l_0,R_0)$ are given by $(0,1.09545)$ for the red curve, $(1.88456,2.43703)$ for the blue curve, and $(2.22055,2.86683)$ for the brown curve.} \label{FigR(l)}
 \end{figure}

Figure~\ref{FigR(l)} shows that, although the geometry remains finite at $l=0$, the location of the bounce depends on the parameter space. For small values of $\bar{M}$, the minimum of the scale function occurs at $l\sim0$, indicating that the bounce remains associated with the short-distance region. As $\bar{M}$ increases, however, the global minimum of $R(l)$ may shift toward a larger finite value of the extended radial coordinate, while $R(0)$ remains finite. The corresponding minimum points $(l_0,R_0)$ are $(0,1.09545)$ for the red curve, $(1.88456,2.43703)$ for the blue curve, and $(2.22055,2.86683)$ for the brown curve.

This behavior does not imply that the curvature-dependent correction becomes relevant in the asymptotic weak-curvature regime, since $F_{\rm eff}\rightarrow0$ at large distances. Rather, the finite behavior at $l=0$ results from the saturation of the effective screened curvature indicator in the high-curvature regime, whereas the location of the bounce is determined by the global profile of the deformed scale function $R(l)$. Thus, depending on the parameters, the proposed construction can affect not only the near-origin geometry but also the global geometric structure of the spacetime. Moreover, it can be directly verified that at the bounce $R'(l_0)=0$ and $R''(l_0)>0$, confirming that these points correspond to genuine minima. For the blue and brown curves, the region around $l\sim0$ behaves as a saddle-like point; nevertheless, the scale function remains finite there with $R'(0)=0$.

We define the mass parameter $M=M_*$ such that the function $A(R(l))=0$ at $l=l_*$
\begin{equation}
   M_*= M(R(l_*))=M(R_*)=\frac{k}{2}R_*+\frac{R_*^3}{2L^2},
\end{equation}
therefore, every value $l=l_*$ represents the location of a horizon in terms of the extended radial coordinate. On the other hand, as discussed above, the scale function $R(l)$ is identified with the Euclidean radial coordinate $r$. Thus, $R(l_*)=R_*$ represents the value at which the scale function, identified with the Euclidean radial coordinate, corresponds to a horizon.

\subsubsection{\bf Spherically symmetric case, $k=1$:}

\begin{figure}[h]   
    \centering 
    \includegraphics[scale=.67]{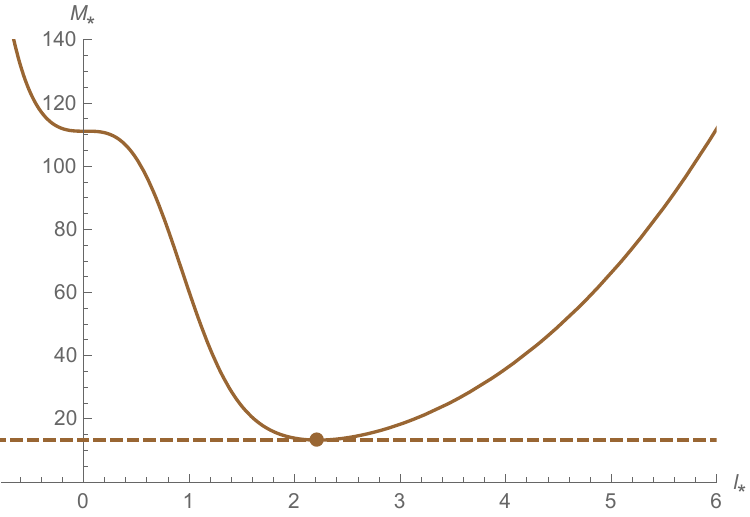} 
    \includegraphics[scale=0.7]{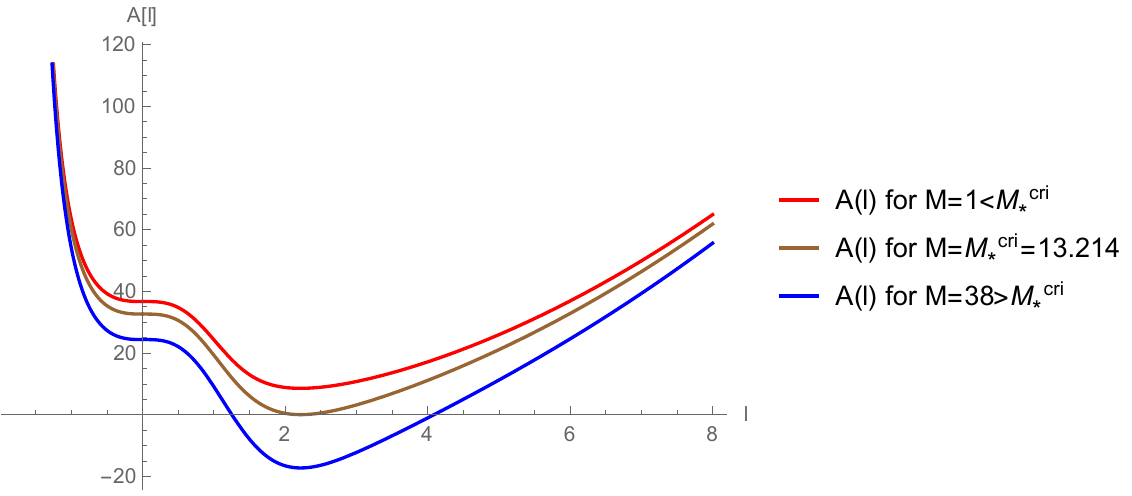} 
    \caption{First panel: The horizontal and vertical axes display $l_*$ and $M_*$, respectively, such that $A(l_*)=0$. The minimum point is given by the ordered pair $(l_*^{(\mathrm{cri})},M_*^{(\mathrm{cri})})=(2.22055,13.2142)$. Second panel: The function $A(l)$ is displayed for $M<M_*^{(\mathrm{cri})}$, $M=M_*^{(\mathrm{cri})}$, and $M>M_*^{(\mathrm{cri})}$ in red, brown, and blue, respectively, representing a traversable wormhole, an extremal RBH, and an RBH. The parameter values used are $a=1,\bar{M}=36,L=1,k=1$.} \label{FigMyFdeLesferico}
 \end{figure}

\begin{figure}[h]   
    \centering 
    \includegraphics[scale=.67]{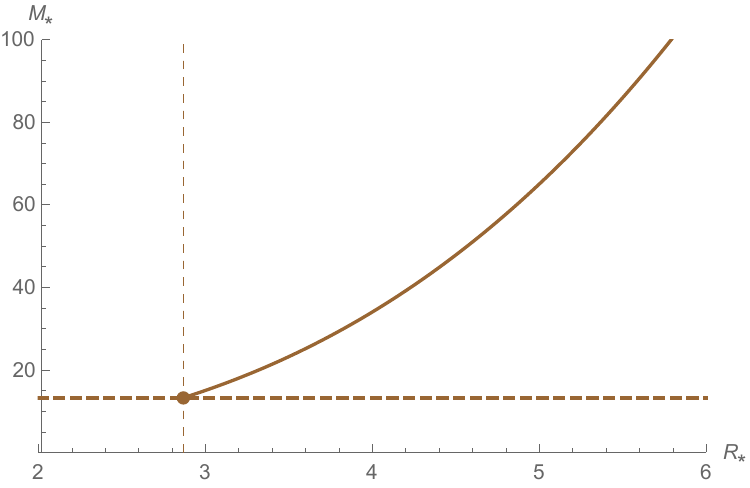} 
    \includegraphics[scale=0.7]{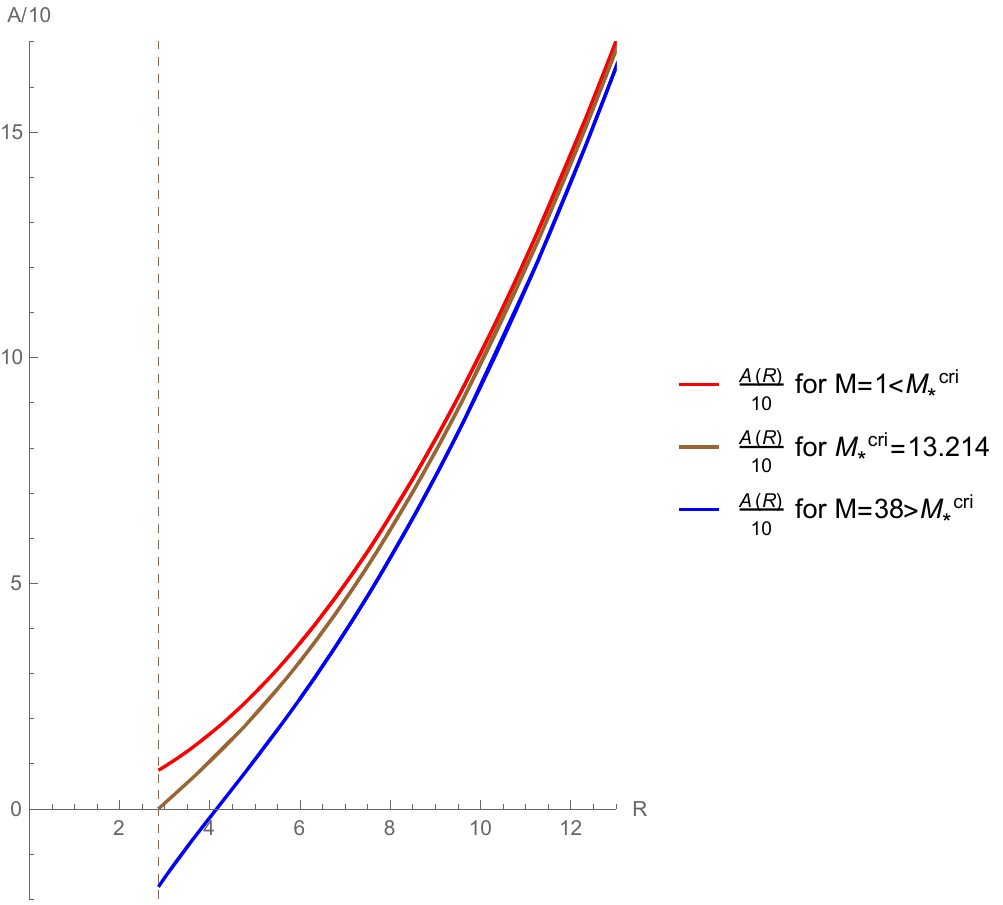} 
    \caption{First panel: The horizontal and vertical axes display $R_*$ and $M_*$, respectively, such that $A(R_*)=0$. The minimum point is given by the ordered pair $(R_*^{(\mathrm{cri})},M_*^{(\mathrm{cri})})=(2.8668,13.2142)$. Second panel: The function $A(R)$ is displayed for $M<M_*^{(\mathrm{cri})}$, $M=M_*^{(\mathrm{cri})}$, and $M>M_*^{(\mathrm{cri})}$ in red, brown, and blue, respectively, representing a traversable wormhole, an extremal RBH, and an RBH. $a=1,\bar{M}=36,L=1,k=1$. The horizon associated with the blue curve is located at $R=R_h=4.1571$. .} \label{FigMyFdeResferico}
 \end{figure}

In order to analyze the behavior with respect to the extended radial coordinate $l$, the first and second panels of Fig. \ref{FigMyFdeLesferico} display the mass parameter $M_*$ as a function of $l_*$, and the function $A(l)$ for different values of the parameter $M$, respectively. Meanwhile, in order to analyze the behavior as a function of the scale function $R$—which can be identified with the Euclidean radial coordinate $r$—Fig. \ref{FigMyFdeResferico} displays the mass parameter $M_*$ and the function $A(R)$ in the first and second panels, respectively. We can point out the following:

\begin{itemize}[leftmargin=5pt]
    \item In the first panel of Fig. \ref{FigMyFdeLesferico} we define $M_*^{(\mathrm{cri})}$ as the minimum value of the mass parameter $M_*$, associated with the ordered pair $(l_*^{(\mathrm{cri})},M_*^{(\mathrm{cri})})$. In this way, we observe that for $M=M_*^{(\mathrm{cri})}$ there exists an extremal RBH whose event horizon is located at $l=l_*^{(\mathrm{cri})}$, which coincides with the point where the bounce occurs in Fig. \ref{FigR(l)}, namely $l_0$. For the parameter values considered in this figure, the ordered pair associated with the minimum is given by $(l_*^{(\mathrm{cri})},M_*^{(\mathrm{cri})})=(2.22055,13.2142)$. We can verify that the bounce, where the minimum of the scale function $R(l)=R(l_0)$ is attained along the brown curve in Fig. \ref{FigR(l)}, satisfies $l_0=l_*^{(\mathrm{cri})}$. It is straightforward to verify that this behavior is generic for other values of the parameters. In the second panel of Fig. \ref{FigMyFdeLesferico}, the brown curve shows that for $M=M_*^{(\mathrm{cri})}$ both horizons coincide, representing an extremal RBH.

    This situation can also be visualized in Fig. \ref{FigMyFdeResferico}. In the first panel, we observe that for $M=M_*^{(\mathrm{cri})}$ there exists an extremal RBH for which $A(R)=0$ at $R=R_0$, where $R_0$ represents the bounce of the scale function identified with the Euclidean radial coordinate. It is worth mentioning that the bounce takes place at the minimum value of $R$, which in the figure corresponds to $R_0=2.8668$ and $M_*^{(\mathrm{cri})}=13.2142$. takes place. In the second panel, we note that, in this case represented by the brown curve, the metric function vanishes exactly at the bounce, and therefore no change of signature. It is worth noting that we can visualize the value $R_0$ corresponding to the bounce in the vertical dashed line in both panels of Figure \ref{FigMyFdeResferico}. 

    \item In the first panel of Fig. \ref{FigMyFdeLesferico}, for $M>M_*^{(\mathrm{cri})}$ there exists an RBH with two horizons satisfying $l_*^{+}>l_*^{-}$. In the second panel of Fig. \ref{FigMyFdeLesferico}, the blue curve shows that for $M>M_*^{(\mathrm{cri})}$ two horizons are present, the largest of which corresponds to the RBH event horizon.

    This situation can also be visualized in Fig. \ref{FigMyFdeResferico}. It can be observed in the first panel that for $M>M_*^{(\mathrm{cri})}$ an event horizon is present. This horizon encloses the value $R_0$, identified with $r_0$, where the bounce occurs, indicating that the geometry corresponds to an RBH. It is worth mentioning that the bounce takes place at the minimum value of $R$, which in the figure corresponds to $R_0=2.8668$ and $M_*^{(\mathrm{cri})}=13.2142$. Moreover, the value $R_0$ associated with the bounce is indicated by the vertical dashed line shown in both panels of Fig. \ref{FigMyFdeResferico}. In the second panel, this scenario is represented by the blue curve, where a change of signature can be observed.

    \item In the first panel of Fig. \ref{FigMyFdeLesferico}, for $M<M_*^{(\mathrm{cri})}$ no horizons are present, and the geometry corresponds to a traversable wormhole. In the second panel of this figure, the red curve shows that for $M<M_*^{(\mathrm{cri})}$ no horizons exist.

    In the first panel of Fig. \ref{FigMyFdeResferico}, for $M<M_*^{(\mathrm{cri})}$ no horizons are present. The minimum value of the scale function is given by $R_0$, identified with $r_0$, implying that no horizon encloses the bounce. This behavior can be visualized in the red curve shown in the second panel. Consequently, the geometry corresponds to a traversable wormhole.
\end{itemize}

\subsubsection{\bf Case with constant hyperbolic transverse section, $k=-1$:}

\begin{figure}[h]   
    \centering 
    \includegraphics[scale=.67]{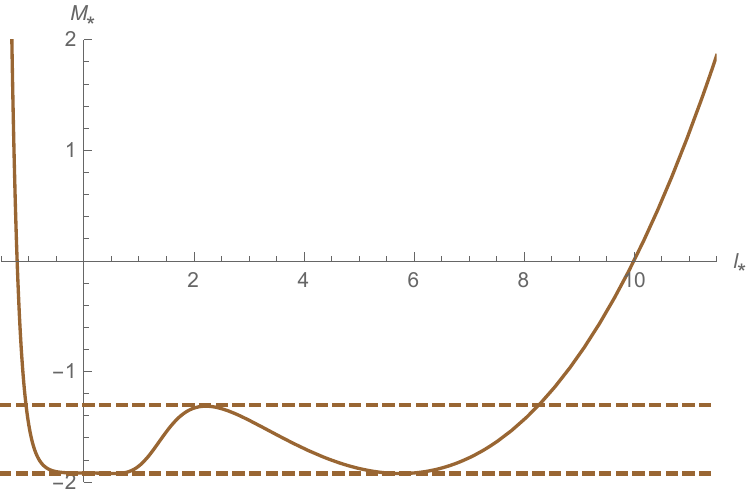} 
    \includegraphics[scale=0.7]{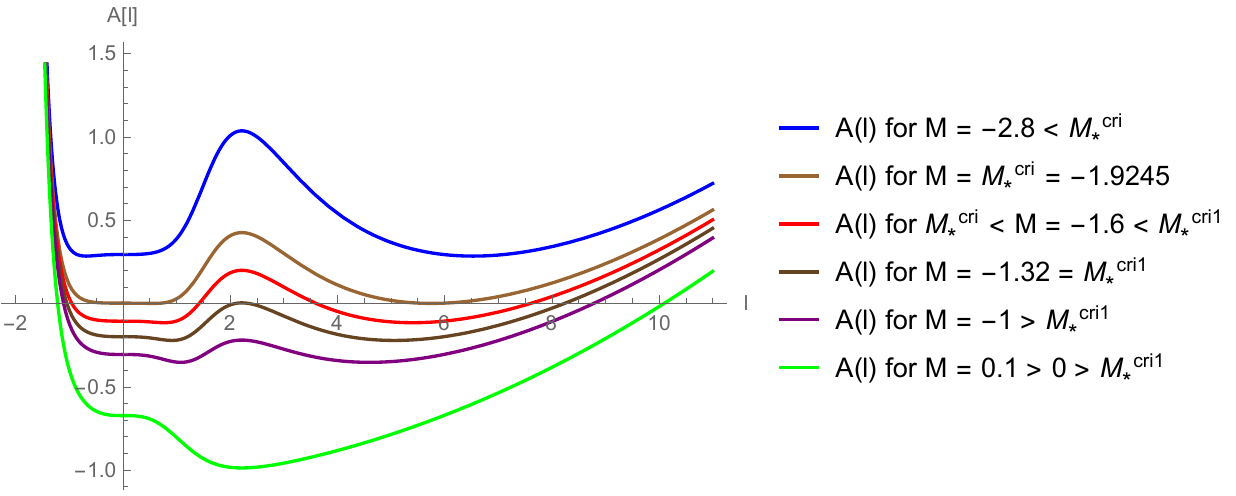} 
    \caption{First panel: Mass parameter $M_*$ as a function of $l_*$ for $a=1$, $\bar{M}=36$, $L=10$, and $k=-1$. Second panel: Function $A(l)$ for $M=-2.8<M_*^{(\mathrm{cri})}$, $M=M_*^{(\mathrm{cri})}=-1.9245$, $M_*^{(\mathrm{cri})}<M=-1.6<M_*^{(\mathrm{cri1})}=-1.32$, $M=M_*^{(\mathrm{cri1})}$, $M=-1>M_*^{(\mathrm{cri1})}<0$, and $M=0.1>M_*^{(\mathrm{cri1})}>0$ with blue, brown, red, brown, purple, and green colors, respectively. } \label{FigMyFdeLhiperbole}
 \end{figure}

\begin{figure}[h]   
    \centering 
    \includegraphics[scale=.67]{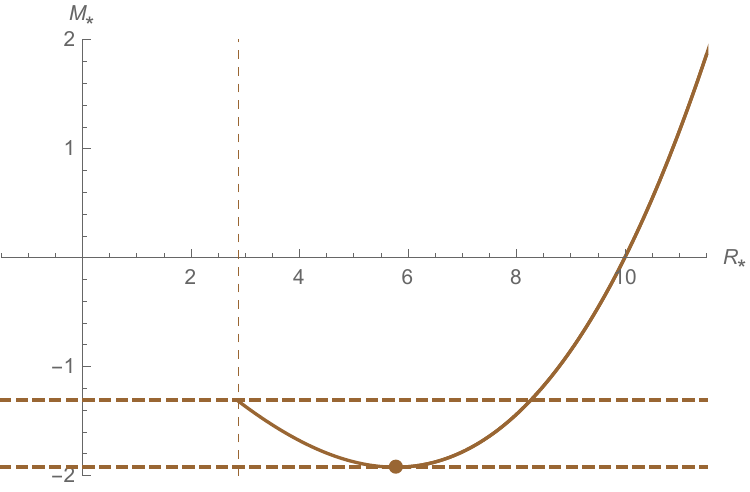} 
    \includegraphics[scale=0.8]{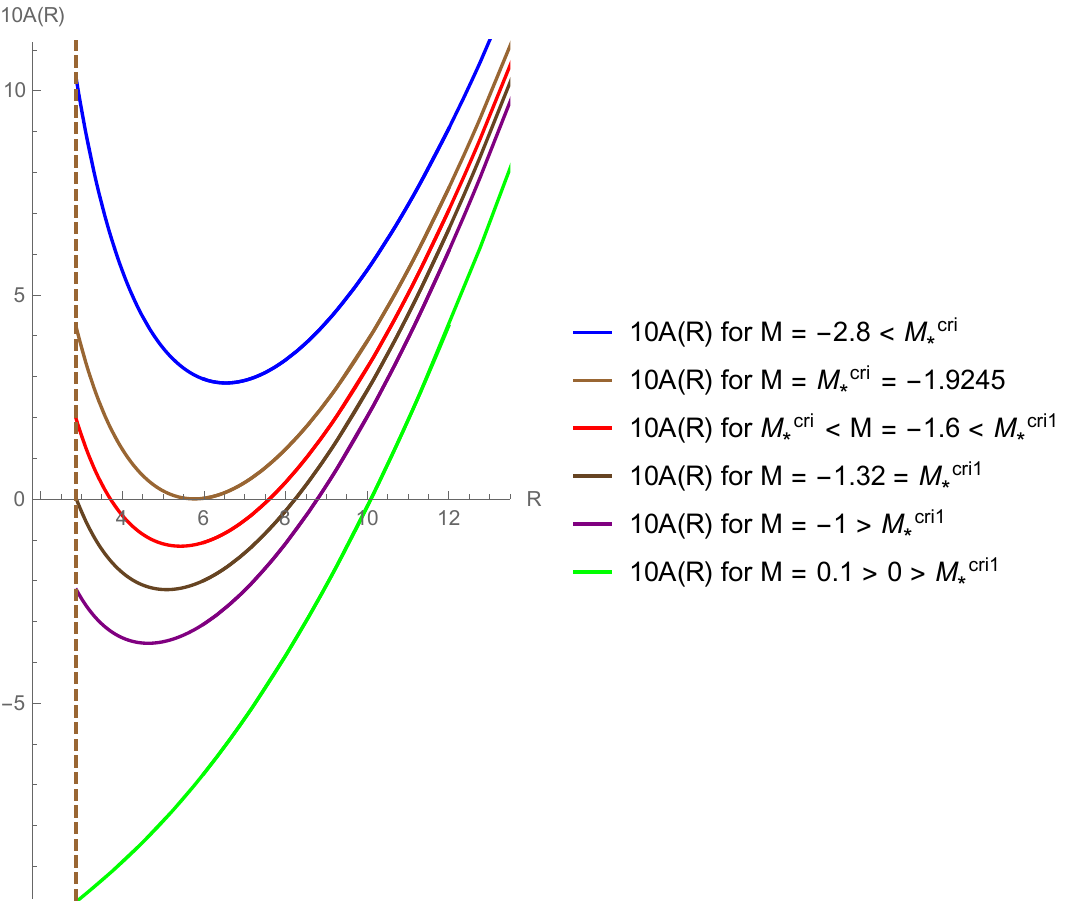} 
    \caption{First panel: Mass parameter $M_*$ as a function of $R_*$ for $a=1$, $\bar{M}=36$, $L=10$, and $k=-1$. Second panel: Function $10 \cdot A(R)$ for $M=-2.8<M_*^{(\mathrm{cri})}$, $M=M_*^{(\mathrm{cri})}=-1.9245$, $M_*^{(\mathrm{cri})}<M=-1.6<M_*^{(\mathrm{cri1})}=-1.32$, $M=M_*^{(\mathrm{cri1})}$, $M=-1>M_*^{(\mathrm{cri1})}<0$, and $M=0.1>M_*^{(\mathrm{cri1})}>0$ with blue, brown, red, brown, purple, and green colors, respectively. } \label{FigMyFdeRhiperbole}
 \end{figure}

As in the spherically symmetric case, the first and second panels of Fig. \ref{FigMyFdeLhiperbole} display the mass parameter $M_*$ as a function of $l_*$, and the function $A(l)$ for different values of the parameter $M$, respectively. Similarly, Fig. \ref{FigMyFdeRhiperbole} displays the mass parameter $M_*$ and the function $A(R)$ as functions of the scale function $R$—identified with the Euclidean radial coordinate $r$—in the first and second panels, respectively. The brown vertical line in the first and second panels represents the minimum value of the scale function, $R_0$, at which the bounce occurs. In the first panel of Fig. \ref{FigMyFdeLhiperbole}, we define $M_*^{(\mathrm{cri})}$ as the minimum value of the mass parameter $M_*$. We also define $M_*^{(\mathrm{cri1})}$ as a local maximum of $M_*$, such that
$M_*^{(\mathrm{cri})}<M_*^{(\mathrm{cri1})}<0$. In this hyperbolic scenario, however, the structure of horizons differs significantly from the spherically symmetric case, as we discuss below:

\begin{itemize}[leftmargin=5pt]

\item We note in the first panel of Fig. \ref{FigMyFdeLhiperbole} that for $M<M_*^{(\mathrm{cri})}$ no horizons are present, and the geometry corresponds to a traversable wormhole admitting negative mass. In the second panel of Fig. \ref{FigMyFdeLhiperbole}, this behavior is represented by the blue curve, where the metric function never vanishes.

This situation can also be visualized in Fig. \ref{FigMyFdeRhiperbole}. In the first panel, for $M<M_*^{(\mathrm{cri})}$ no horizons are present, implying that no horizon encloses the bounce located at $R_0$, identified with $r_0$. In the second panel, represented by the blue curve, the metric function does not vanish. Consequently, the geometry corresponds to a traversable wormhole with negative mass.

\item In the first panel of Fig. \ref{FigMyFdeLhiperbole}, for $M=M_*^{(\mathrm{cri})}$ there exists an extremal configuration for which two horizons share the same value $M=M_*^{(\mathrm{cri})}$. In the second panel of Fig. \ref{FigMyFdeLhiperbole}, this behavior is represented by the brown curve, where the metric function vanishes twice without any change of signature.

This situation can also be visualized in Fig. \ref{FigMyFdeRhiperbole}. In the first panel, we observe that for $M=M_*^{(\mathrm{cri})}$ there exists an extremal RBH for which the function $A(R)$ vanishes at a single value, enclosing the bounce located at $R=R_0$. In the second panel, represented by the brown curve, the metric function touches the horizontal axis once without any change of signature, corresponding to an extremal RBH configuration.

\item In the first panel of Fig. \ref{FigMyFdeLhiperbole}, for $M_*^{(\mathrm{cri})}<M<M_*^{(\mathrm{cri1})}$ there exist four horizons with respect to the extended radial coordinate $l$. Therefore, the geometry corresponds to an RBH configuration. In the second panel of Fig. \ref{FigMyFdeLhiperbole}, this behavior is represented by the red curve, where four zeros of the metric function are present.

This situation can also be visualized in Fig. \ref{FigMyFdeRhiperbole}. In the first panel, for $M_*^{(\mathrm{cri})}<M<M_*^{(\mathrm{cri1})}$ there exist two horizons located at $R_+$ and $R_-$, where $R_+$ corresponds to the event horizon and $R_-$ to the inner horizon. Both the event and inner horizons enclose the bounce located at $R_0$, indicating that the geometry corresponds to an RBH. In the second panel, represented by the red curve, a change of signature takes place between both horizons. It is worth mentioning that this RBH configuration admits negative mass.

\item In the first panel of Fig. \ref{FigMyFdeLhiperbole}, for $M_*=M_*^{(\mathrm{cri1})}$ there exist three horizons with respect to the extended radial coordinate $l$, where the second one corresponds to a local maximum. Therefore, the geometry corresponds to an RBH configuration. In the second panel of Fig. \ref{FigMyFdeLhiperbole}, this behavior is represented by the lower brown curve, where three zeros of the metric function are present.

This situation can also be visualized in Fig. \ref{FigMyFdeRhiperbole}. In the first panel, for $M_*=M_*^{(\mathrm{cri1})}$ there exist two horizons located at $R_+$ and $R_-$, where $R_+$ corresponds to the event horizon and $R_-$ to the inner horizon. The inner horizon coincides with the location of the bounce. Both the event and inner horizons enclose the bounce located at $R_0$, indicating that the geometry corresponds to an RBH. In the second panel, represented by the lower brown curve, a change of signature takes place between both horizons. It can also be observed that the inner horizon coincides with the location of the bounce. It is worth mentioning that this RBH configuration admits negative mass.

\item In the first panel of Fig. \ref{FigMyFdeLhiperbole}, for $M>M_*^{(\mathrm{cri1})}$ there exist two horizons with respect to the extended radial coordinate $l$, corresponding to an RBH geometry. This behavior occurs for both negative and positive values of the mass parameter $M$. In the second panel of Fig. \ref{FigMyFdeLhiperbole}, this situation is represented by the green and purple curves, where two zeros of the metric function are present.

This behavior can also be visualized in Fig. \ref{FigMyFdeRhiperbole}. In the first panel, for $M>M_*^{(\mathrm{cri1})}$ the geometry possesses a single event horizon located at $R_+$, which encloses the bounce at $R_0$. Therefore, the geometry corresponds to an RBH for both negative and positive mass configurations. In the second panel, represented by the green and purple curves, the metric function vanishes at a single value of $R$, corresponding to the event horizon.

\end{itemize}

\subsubsection{\bf Case with constant planar transverse section, $k=0$:}

\begin{figure}[h]   
    \centering 
    \includegraphics[scale=.67]{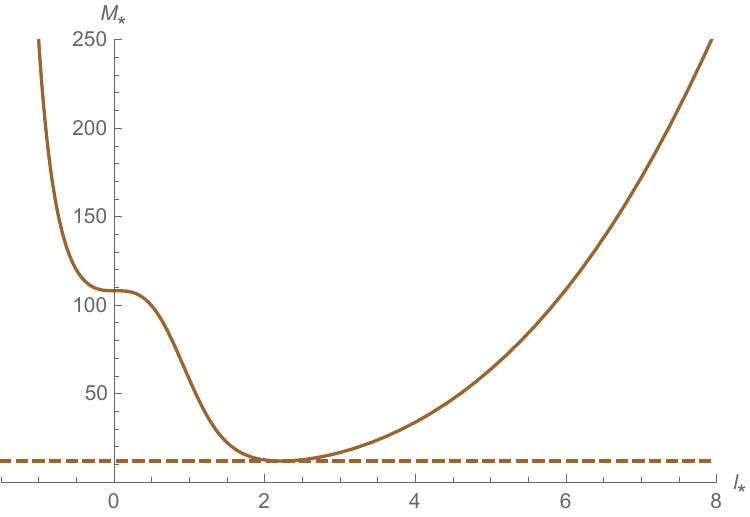} 
    \includegraphics[scale=0.7]{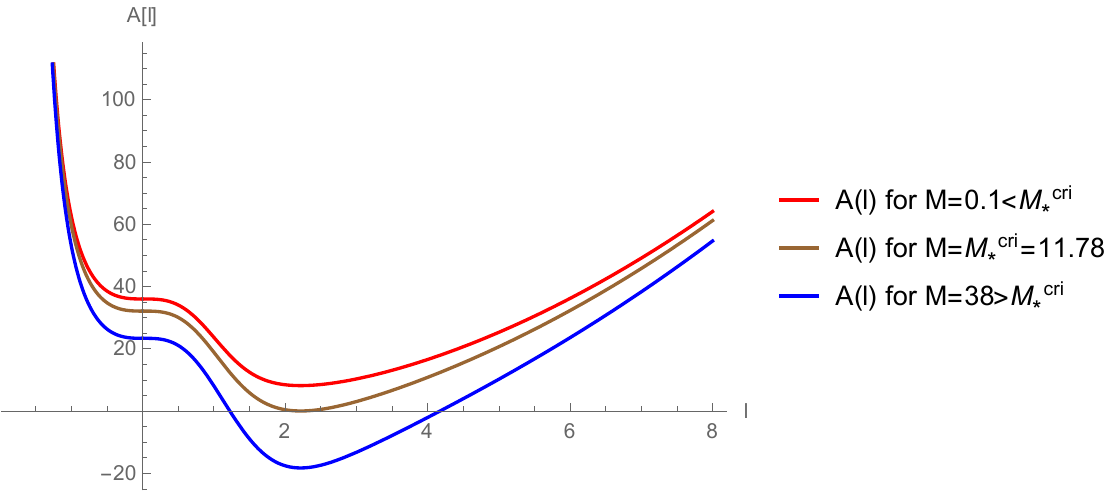} 
    \caption{First panel: The horizontal and vertical axes display $l_*$ and $M_*$, respectively, such that $A(l_*)=0$. The minimum point is given by the ordered pair $(l_*^{(\mathrm{cri})},M_*^{(\mathrm{cri})})=(2.22055,11.7808)$. Second panel: The function $A(l)$ is displayed for $M<M_*^{(\mathrm{cri})}$, $M=M_*^{(\mathrm{cri})}$, and $M>M_*^{(\mathrm{cri})}$ in red, brown, and blue, respectively, representing a traversable wormhole, an extremal RBH, and an RBH. The parameter values used are $a=1,\bar{M}=36,L=1,k=0$.} \label{FigMyFdeLplanar}
 \end{figure}

\begin{figure}[h]   
    \centering 
    \includegraphics[scale=.67]{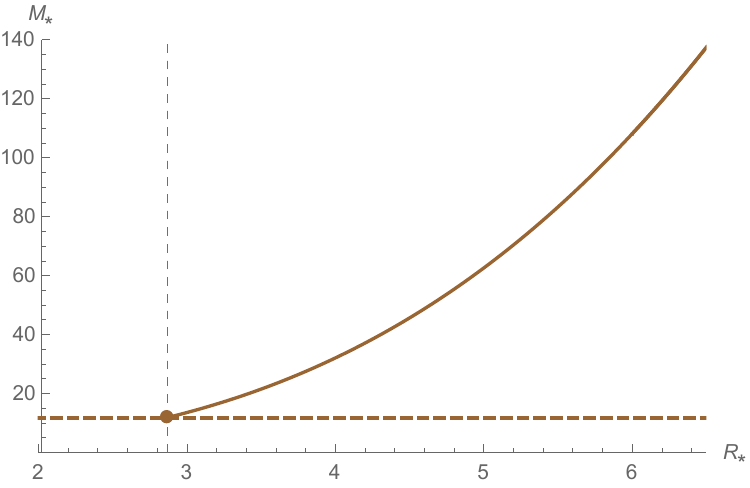} 
    \includegraphics[scale=0.7]{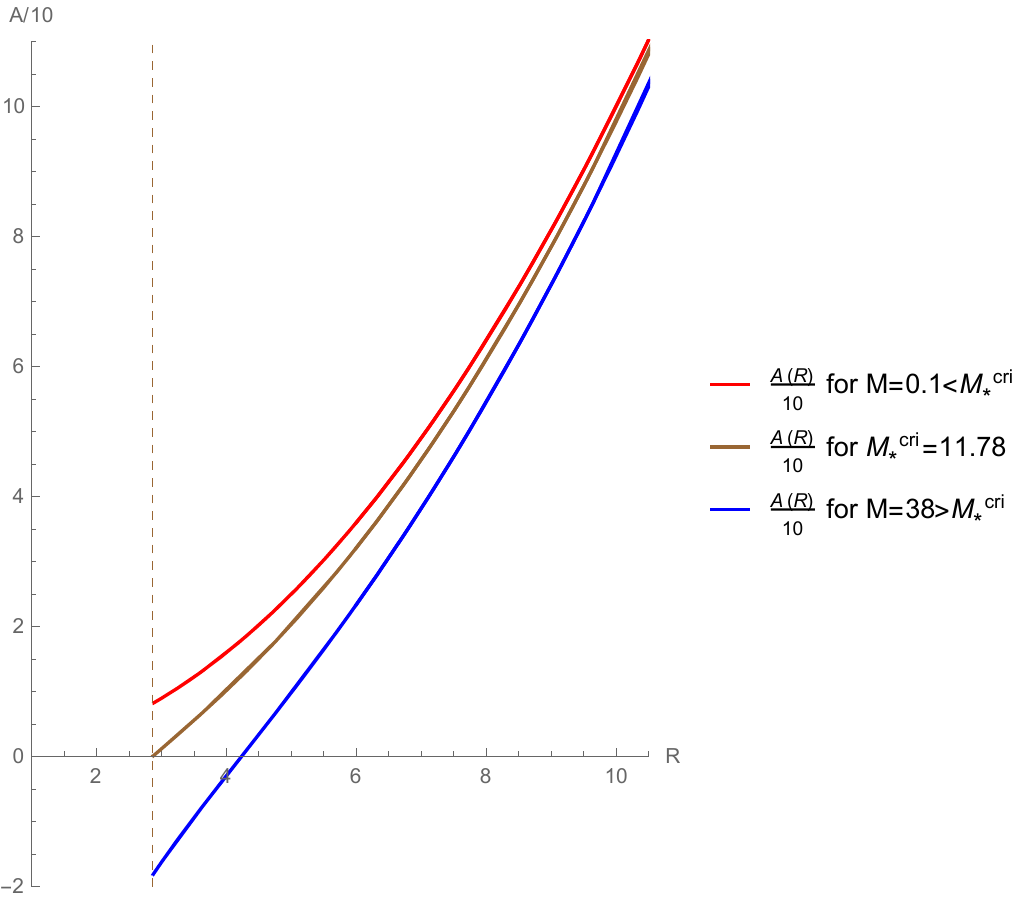} 
    \caption{First panel: The horizontal and vertical axes display $R_*$ and $M_*$, respectively, such that $A(R_*)=0$. The minimum point is given by the ordered pair $(R_*^{(\mathrm{cri})},M_*^{(\mathrm{cri})})=(2.8668,11.7808)$. Second panel: The function $A(R)$ is displayed for $M<M_*^{(\mathrm{cri})}$, $M=M_*^{(\mathrm{cri})}$, and $M>M_*^{(\mathrm{cri})}$ in red, brown, and blue, respectively, representing a traversable wormhole, an extremal RBH, and an RBH. The parameter values used are $a=1,\bar{M}=36,L=1,k=0$. The horizon associated with the blue curve is located at $R=R_h=4.17711$.} \label{FigMyFdeRplanar}
 \end{figure}

In order to analyze the behavior with respect to the extended radial coordinate $l$, the first and second panels of Fig. \ref{FigMyFdeLplanar} display the mass parameter $M_*$ as a function of $l_*$, and the function $A(l)$ for different values of the parameter $M$, respectively. Meanwhile, in order to analyze the behavior as a function of the scale function $R$—which can be identified with the Euclidean radial coordinate $r$—Fig. \ref{FigMyFdeRplanar} displays the mass parameter $M_*$ and the function $A(R)$ in the first and second panels, respectively. We can point out the following:

\begin{itemize}[leftmargin=5pt]

    \item In the first panel of Fig. \ref{FigMyFdeLplanar} we define $M_*^{(\mathrm{cri})}$ as the minimum value of the mass parameter $M_*$, associated with the ordered pair $(l_*^{(\mathrm{cri})},M_*^{(\mathrm{cri})})$. In this way, we observe that for $M=M_*^{(\mathrm{cri})}$ there exists an extremal RBH whose event horizon is located at $l=l_*^{(\mathrm{cri})}$, which coincides with the point where the bounce occurs in Fig. \ref{FigR(l)}, namely $l_0$. For the parameter values considered in this figure, the ordered pair associated with the minimum is given by $(l_*^{(\mathrm{cri})},M_*^{(\mathrm{cri})})=(2.22055,11.7808)$. We can verify that the bounce, where the minimum of the scale function $R(l)=R(l_0)$ is attained along the brown curve in Fig. \ref{FigR(l)}, satisfies $l_0=l_*^{(\mathrm{cri})}$. It is straightforward to verify that this behavior is generic for other values of the parameters. In the second panel of Fig. \ref{FigMyFdeLplanar}, the brown curve shows that for $M=M_*^{(\mathrm{cri})}$ both horizons coincide, representing an extremal RBH.

    This situation can also be visualized in Fig. \ref{FigMyFdeRplanar}. In the first panel, we observe that for $M=M_*^{(\mathrm{cri})}$ there exists an extremal RBH for which $A(R)=0$ at $R=R_0$, where $R_0$ represents the bounce of the scale function identified with the Euclidean radial coordinate. It is worth mentioning that the bounce takes place at the minimum value of $R$, which in the figure corresponds to $R_0=2.8668$ and $M_*^{(\mathrm{cri})}=11.7808$. In the second panel, we note that, in this case represented by the brown curve, the metric function vanishes exactly at the bounce, and therefore no change of signature occurs. It is worth noting that we can visualize the value $R_0$ corresponding to the bounce in the vertical dashed line in both panels of Fig. \ref{FigMyFdeRplanar}.

    \item In the first panel of Fig. \ref{FigMyFdeLplanar}, for $M>M_*^{(\mathrm{cri})}$ there exists an RBH with two horizons satisfying $l_*^{+}>l_*^{-}$. In the second panel of Fig. \ref{FigMyFdeLplanar}, the blue curve shows that for $M>M_*^{(\mathrm{cri})}$ two horizons are present, the largest of which corresponds to the RBH event horizon.

    This situation can also be visualized in Fig. \ref{FigMyFdeRplanar}. It can be observed in the first panel that for $M>M_*^{(\mathrm{cri})}$ an event horizon is present. This horizon encloses the value $R_0$, identified with $r_0$, where the bounce occurs, indicating that the geometry corresponds to an RBH. It is worth mentioning that the bounce takes place at the minimum value of $R$, which in the figure corresponds to $R_0=2.8668$ and $M_*^{(\mathrm{cri})}=11.7808$. Moreover, the value $R_0$ associated with the bounce is indicated by the vertical dashed line shown in both panels of Fig. \ref{FigMyFdeRplanar}. In the second panel, this scenario is represented by the blue curve, where a change of signature can be observed.

    \item In the first panel of Fig. \ref{FigMyFdeLplanar}, for $M<M_*^{(\mathrm{cri})}$ no horizons are present, and the geometry corresponds to a traversable wormhole. In the second panel of this figure, the red curve shows that for $M<M_*^{(\mathrm{cri})}$ no horizons exist.

    In the first panel of Fig. \ref{FigMyFdeRplanar}, for $M<M_*^{(\mathrm{cri})}$ no horizons are present. The minimum value of the scale function is given by $R_0$, identified with $r_0$, implying that no horizon encloses the bounce. This behavior can be visualized in the red curve shown in the second panel. Consequently, the geometry corresponds to a traversable wormhole.

\end{itemize}

\subsection{The Kretschmann scalar}

It is also worth testing the behavior of the Kretschmann scalar for the three previously discussed cases: spherical, hyperbolic, and planar constant-transverse-section geometries, corresponding to $k=1$, $k=-1$, and $k=0$, respectively. The value of the latter for our space--time is given by:

\begin{equation}
    K= \big (A''(l) \big)^2+ 2 \left ( \frac{A'(l)\,R'(l)}{R(l)} \right)^2+4 \left ( \frac{-k+A(l)\,R'(l)^2}{R(l)^2}\right )^2
    + 2 \left ( \frac{A'(l)\,R'(l)+ 2A(l)\, R''(l)}{R(l)} \right )^2
\end{equation}

From the previous analysis, we can observe that in all three cases considered—namely, spherical, hyperbolic, and planar transverse sections—the scale function possesses a nonvanishing minimum, $R(l)>0$, at which the bounce occurs. Consequently, the Kretschmann scalar does not exhibit divergences associated with the vanishing of this function. Moreover, the graphical analysis indicates that neither the scale function nor $A(l)$, together with their derivatives, develop divergences. Therefore, the Kretschmann scalar provides information that the geometry remains regular everywhere.

\subsection{Sources and Energy Conditions}

We consider the following form for the energy–momentum tensor:
\begin{equation} \label{TensorEM}
T^\mu_\nu = \mathrm{diag} \Big (-\rho(l), p_r (l), p_{\theta}(l), p_{\theta}(l)\Big)
\end{equation}
whose components are
\begin{align}
    &\frac{1}{8\pi\,R(l)^2}\,\Big ( -k + A'(l)\,R'(l)\,R(l)+A(l)\,R'(l)^2+2\,A(l)\,R(l)\,R''(l) \Big )-\frac{3}{8\pi\,L^2} =-\rho(l) \label{densidad} \\
    &\frac{1}{8\pi\,R(l)^2}\,\Big ( -k + A'(l)\,R'(l)\,R(l)+A(l)\,R'(l)^2 \Big )-\frac{3}{8\pi\,L^2} =p_r (l) \label{presionRadial} \\
    &\frac{1}{8\pi\,R(l)}\,\Big ( A'(l)\,R'(l)+\frac{1}{2}A''(l)\,R(l)+A(l)\,R''(l) \Big )-\frac{3}{8\pi\,L^2} =p_\theta (l) \label{presionTangencial}
\end{align}
where $'$ denotes differentiation with respect to the coordinate $l$ and where $k=1,-1,0$ when the transverse section in the last equation corresponds, respectively, to a sphere $\mathcal{S}^2$, a hyperboloid $\mathcal{H}^2$, or a plane $\mathcal{R}^2$, described by the transverse line element $ds_{T-k}^2$ in Eq. \eqref{ElementoTransversal}. The scale function $R(l)$ is given by Eqs. \eqref{FefectivoReemplazado} and \eqref{FuncionEscala1}, while the function $A(l)$ is determined by Eq. \eqref{FuncionA}. The conservation equation, through $\nabla_\mu T^{\mu \nu}=0$, is given by:
\begin{equation}
    \frac{A'(l)}{2A(l)} \Big ( \rho(l)+ p_r (l)\Big) +p_r' (l)+\frac{2R'(l)}{R(l)} \Big (  p_r (l)-p_\theta (l)\Big) =0
\end{equation}
where, as is well known, the Bianchi identities imply that the system consists of four equations, of which only three are independent. In this way, the components of the energy--momentum tensor are obtained directly by substituting the corresponding value of $k$ into $A(l)$ and $R(l)$ in Eqs. \eqref{densidad}, \eqref{presionRadial}, \eqref{presionTangencial}.

For simplicity, we analyze the behavior of $\rho$, $\rho+p_r$, and $\rho+p_r+2p_\theta$ with respect to the scale function $R$, which, as discussed above, is identified with the Euclidean radial coordinate $r$. These quantities are used to test the Weak, Null, and Strong Energy Conditions (WEC, NEC, and SEC), respectively. This procedure allows us to examine these energy conditions more explicitly in the vicinity of the bounce, where $R$ reaches its minimum value. The procedure is repeated for the three cases considered in this work, namely, geometries with spherical, planar, and hyperbolic transverse sections, corresponding to $k=1$, $k=0$, and $k=-1$, respectively. In all cases, the parameter values have been chosen in accordance with the analyses presented above.

\begin{figure}[p]   
    \centering 
    \includegraphics[scale=.55]{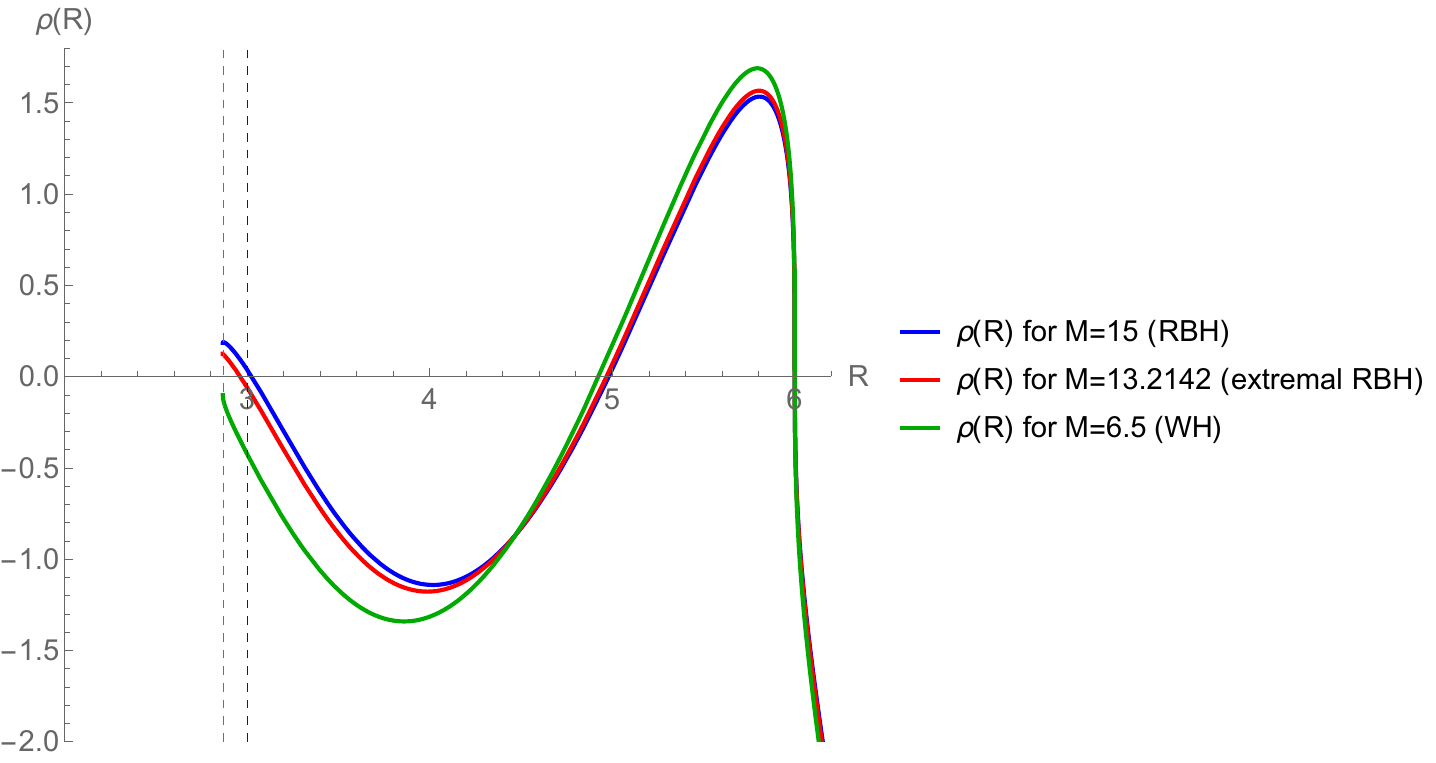} 
    \includegraphics[scale=0.55]{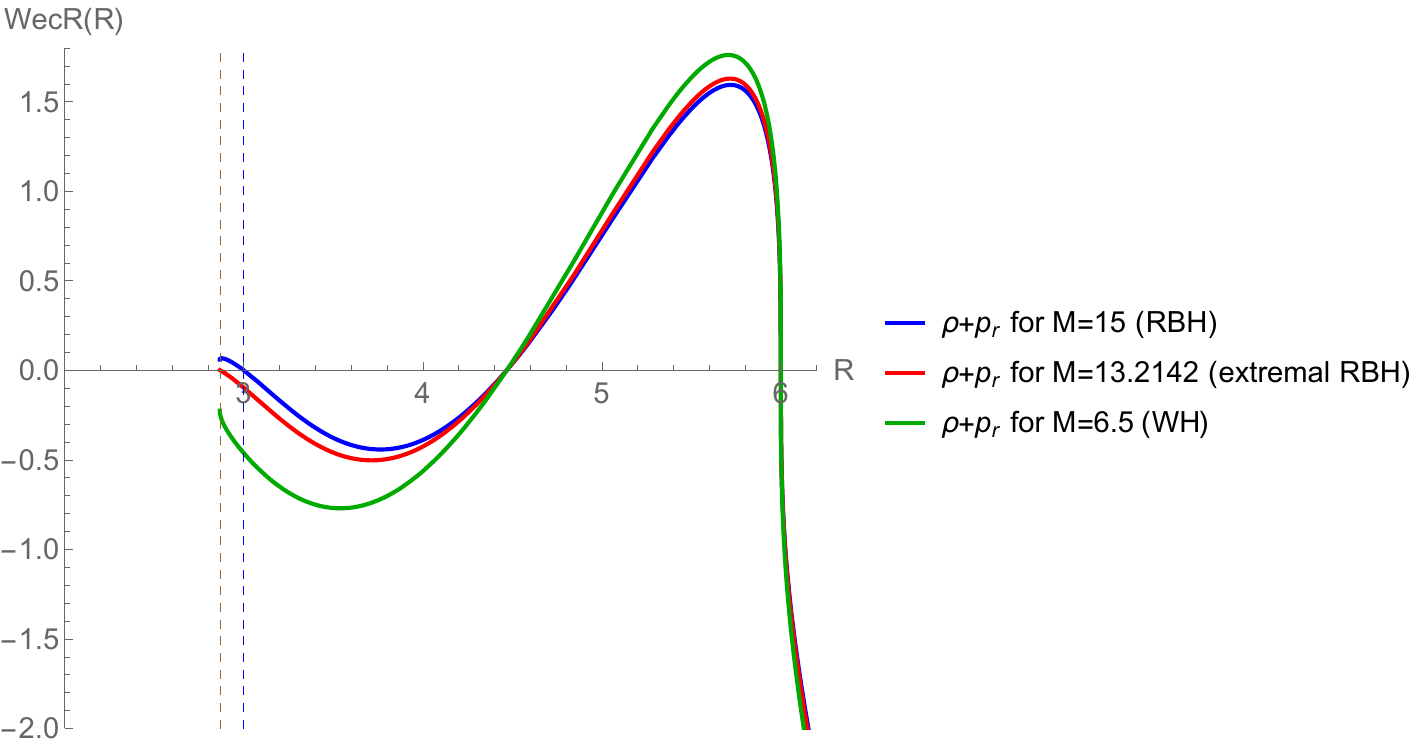} 
    \includegraphics[scale=0.7]{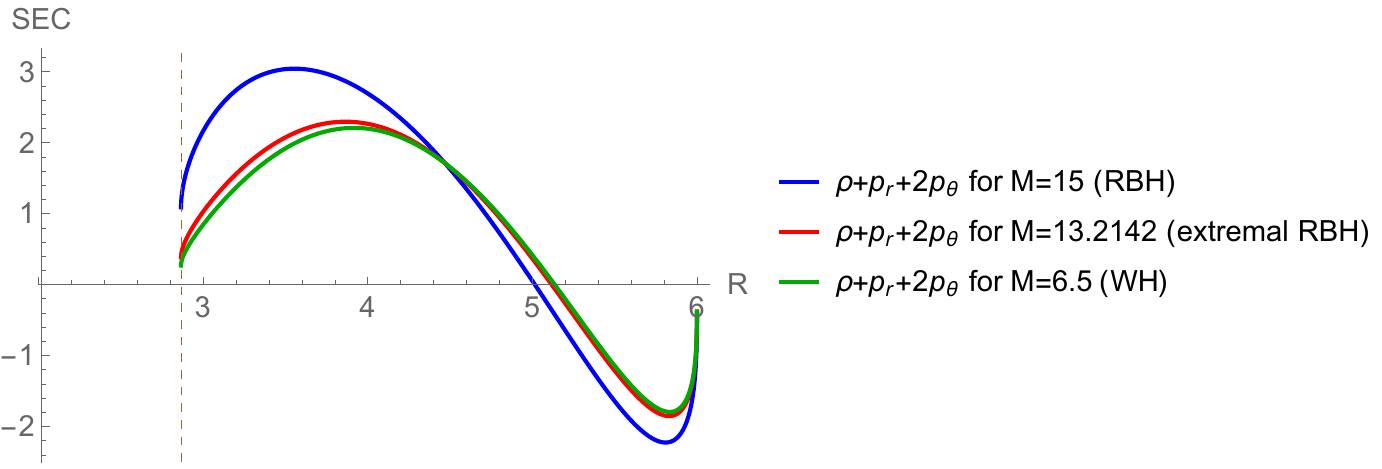} 
    \caption{For $a=1$, $\bar{M}=36$, $L=1$, and $k=1$, we consider an RBH with $M=15$, an extremal RBH with $M=13.2142$, and a WH with $M=6.5$. The horizontal axis represents $R$. The vertical axis represents $\rho$, $\rho+p_r$, and $\rho+p_r+2p_\theta$ in the first, second, and third panels, respectively. The brown and blue dashed  vertical lines indicate the locations of the bounce and the event horizon of the RBH, respectively. } \label{FigCondicionesEnergiaEsferico}
 \end{figure}

In Fig.~\ref{FigCondicionesEnergiaEsferico}, for the spherical case with $k=1$, the horizontal axis corresponds to the scale function $R$, while the vertical axis represents $\rho$, $\rho+p_r$, and $\rho+p_r+2p_\theta$ in the first, second, and third panels, respectively. The parameters shown in the figure have been chosen in accordance with the analysis carried out in the previous sections.
The brown dashed vertical line indicates the location of the bounce, which, in the wormhole case, coincides with the throat. We find that, for both the RBH and the extremal Regular Black hole (eRBH), $\rho>0$ in the vicinity of the bounce. The blue dashed line marks the location of the event horizon of the RBH at $R=3$, showing that the WEC is satisfied at the event horizon. In both the RBH and eRBH cases, as $R$ increases, the energy density becomes negative, violating the WEC, and later returns to positive values at finite distances. In contrast, for the wormhole (WH) geometry, $\rho<0$ near the bounce, which corresponds to the throat, indicating a violation of the WEC in this region, as expected. As $R$ increases further, the energy density becomes positive before turning negative again at larger distances.

The second panel displays the behavior of $\rho+p_r$. Interestingly, for the eRBH, one finds that $\rho=-p_r$ at the location of the bounce, as occurs in the usual spacetimes whose angular sector is given by $r^2 d\Omega^2$. For both the RBH and eRBH, the NEC is satisfied near the bounce, then violated, and finally restored at larger scales. In the wormhole case, the NEC is violated in the vicinity of the throat and again at large distances, as expected. Finally, the third panel tests the SEC through the quantity $\rho+p_r+2p_\theta$. We find that the SEC is satisfied near the bounce for all three geometries: the RBH, the eRBH, and the wormhole.

\begin{figure}[h]   
    \centering 
    \includegraphics[scale=.55]{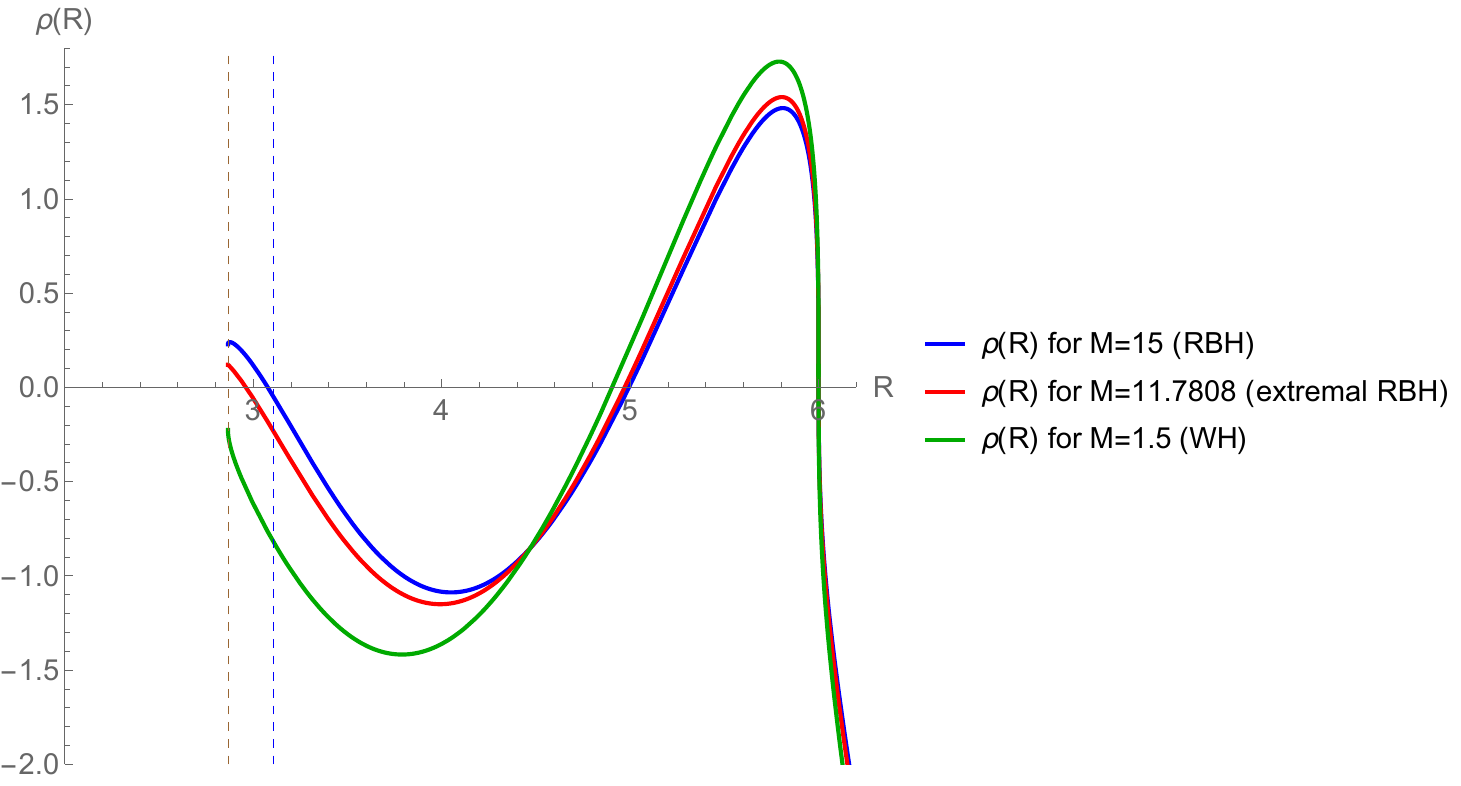} 
    \includegraphics[scale=0.55]{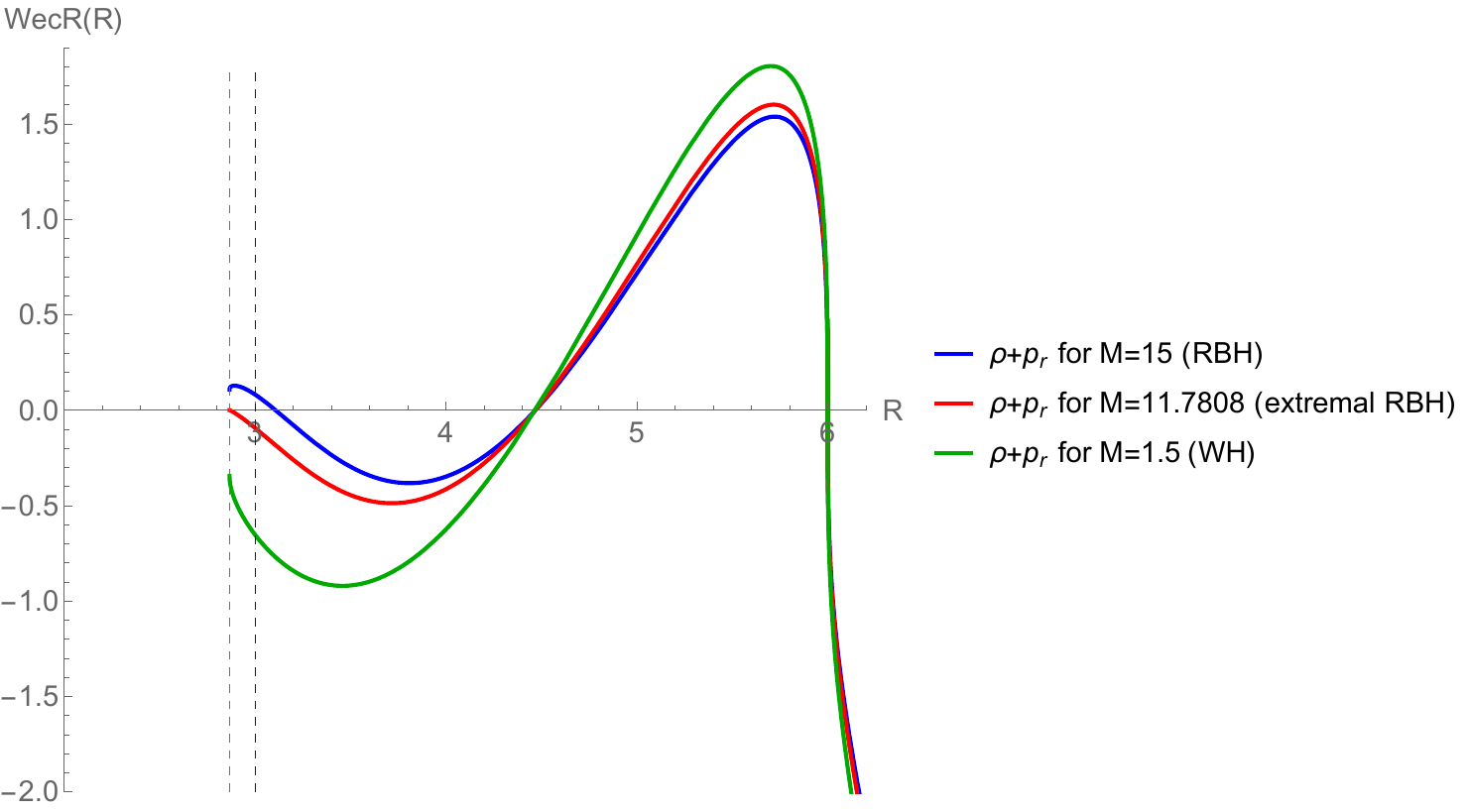} 
    \includegraphics[scale=0.7]{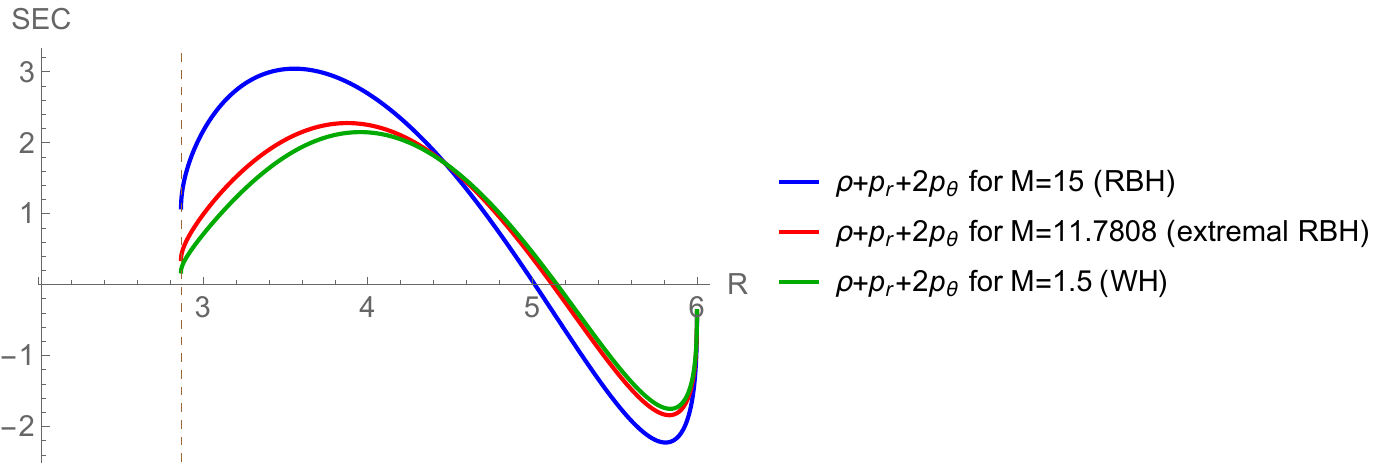} 
    \caption{For $a=1$, $\bar{M}=36$, $L=1$, and $k=0$, we consider an RBH with $M=55$, an extremal RBH with $M=11.7808$, and a WH with $M=1.5$. The horizontal axis represents $R$. The vertical axis represents $\rho$, $\rho+p_r$, and $\rho+p_r+2p_\theta$ in the first, second, and third panels, respectively. The brown and blue dashed  vertical lines indicate the locations of the bounce and the event horizon of the RBH, respectively. } \label{FigCondicionesEnergiaPlanar}
 \end{figure}

An analogous procedure has been followed in the first, second, and third panels of Fig. \ref{FigCondicionesEnergiaPlanar} for the planar case with $k=0$. We note that, for the parameter values chosen in this figure, the behavior of the energy conditions is analogous to that found in the spherical case.

\begin{figure}[h]   
    \centering 
    \includegraphics[scale=.7]{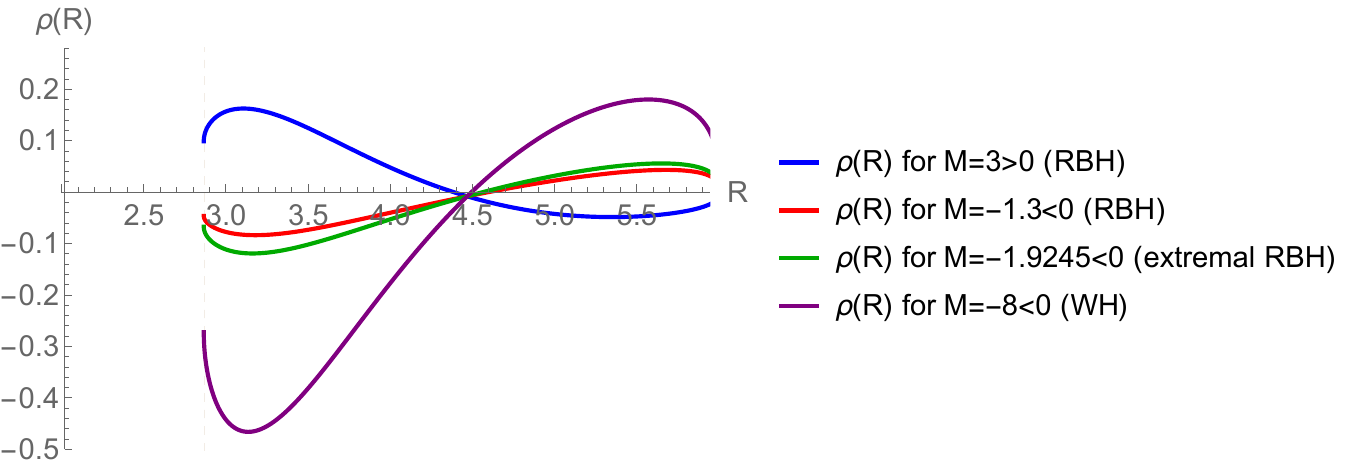} 
    \includegraphics[scale=0.7]{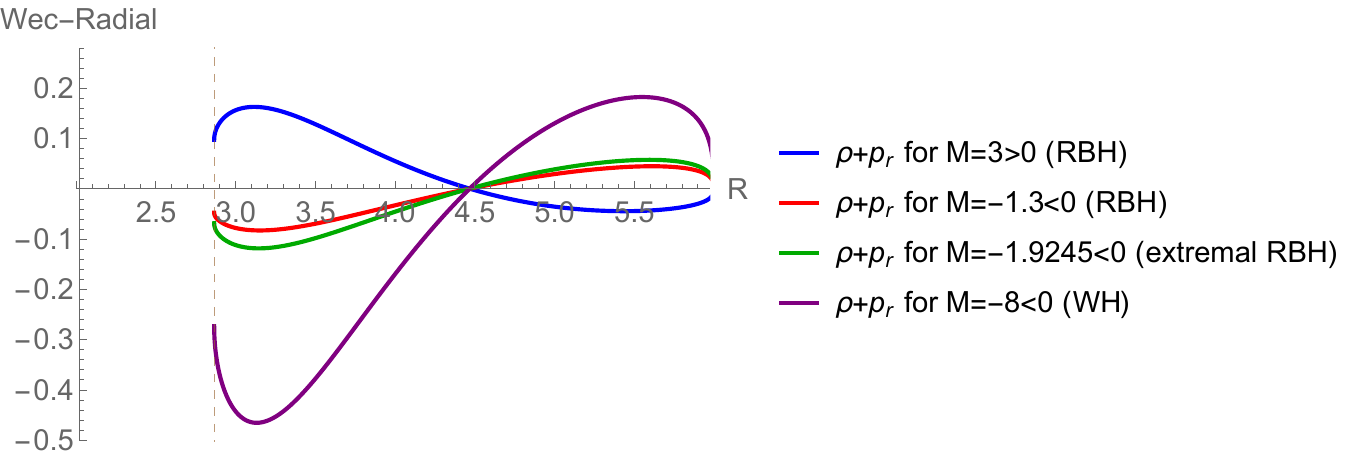} 
    \includegraphics[scale=0.7]{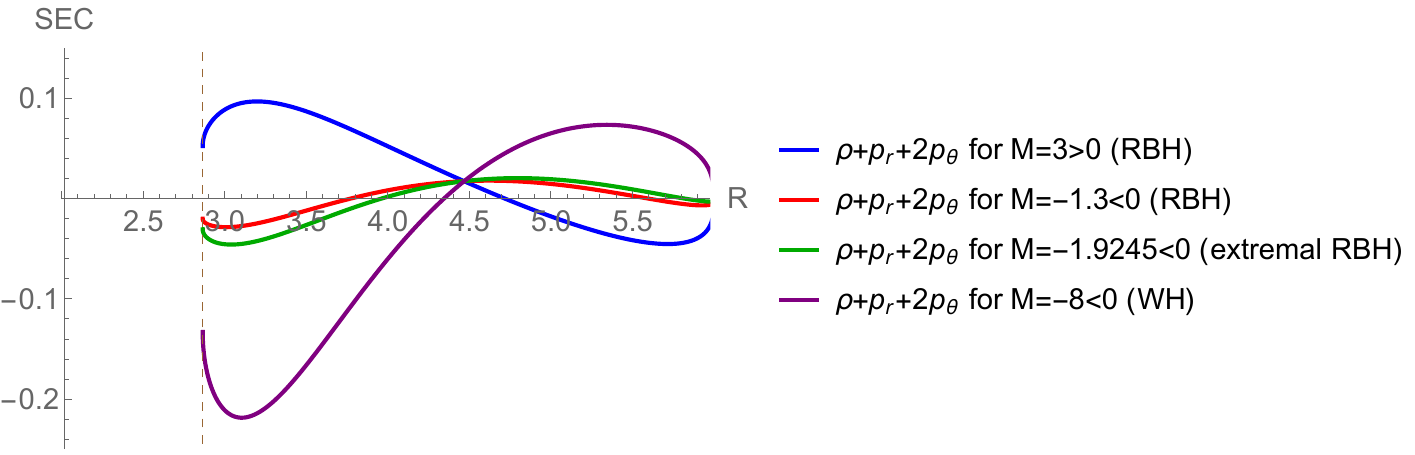} 
    \caption{For $a=1$, $\bar{M}=36$, $L=10$, and $k=-1$, we consider an RBH with $M=3>0$ and $M=-1.3<0$, an extremal RBH with $M=-1.9245$, and a WH with $M=-8$. The horizontal axis represents $R$. The vertical axis represents $\rho$, $\rho+p_r$, and $\rho+p_r+2p_\theta$ in the first, second, and third panels, respectively. The brown dashed vertical lines indicate the locations of the bounce. } \label{FigCondicionesEnergiaHiperbole}
 \end{figure}

From Fig.~\ref{FigCondicionesEnergiaHiperbole}, we can verify that, for the hyperbolic case, the behavior of the energy conditions differs from that found in the spherical and planar cases. First, we note that, in the wormhole case, as expected, $\rho<0$ and $\rho+p_r<0$ near the throat. Moreover, the DEC is also violated in the vicinity of this location. For the extremal RBH, the aforementioned energy conditions are likewise violated near the bounce. The same behavior is observed for the RBH with negative mass. In contrast, for the positive-mass RBH, all these energy conditions are satisfied in the vicinity of the bounce.

Physically, these results show that the presence of a geometric bounce does not necessarily imply the violation of the energy conditions in its vicinity in the case of regular black holes. In particular, for the spherical and planar cases, the regular black hole geometries satisfy the WEC, NEC, and SEC near the bounce, suggesting that singularity regularization can be achieved through non-exotic matter sources in this region. In contrast, for wormhole geometries, the violation of the WEC and NEC near the throat remains a generic feature, in agreement with the classical results for traversable wormholes. Moreover, the hyperbolic case exhibits a more intriguing behavior, since the fulfillment of the energy conditions depends sensitively on the value of the mass parameter. In this way, the topology of the transverse section plays an important role in determining the physical nature of the matter sources supporting these regular geometries.

\section{Conclusion and discussion}

In this work, we have developed an effective curvature-dependent construction for black-bounce spacetimes in which the constant minimal-length contribution of the original Simpson--Visser model is replaced by an effective screened curvature indicator carrying information about the corresponding vacuum geometry. The construction is motivated by the Kretschmann invariant as a natural indicator of the high-curvature regime and by curvature-dependent screening prescriptions previously employed in regular black-hole geometries. In particular, the divergent behavior of the vacuum curvature indicator $F\sim K_v^{1/2}$ is mapped, through the phenomenological screening function, into the finite quantity $F_{\rm eff}$, which is subsequently incorporated into the scale function. In this way, the deformation becomes relevant in the region where the curvature of the vacuum counterpart becomes unbounded, while becoming almost negligible asymptotically. The resulting geometry therefore preserves the asymptotic vacuum behavior while remaining finite at short distances.

A distinctive consequence of this construction is that the saturation of $F_{\rm eff}$ ensures the finiteness of the deformed geometry at the origin of the extended radial coordinate, $l=0$, whereas the location of the bounce is determined by the complete profile of the resulting scale function $R(l)$. Depending on the parameter space, its minimum may remain in the short-distance region or shift toward a larger finite value of $l$, while the geometry continues to be finite at the origin. Therefore, the short-distance saturation and the location of the bounce represent two related but distinct features of the construction: the former controls the regular behavior near the origin, whereas the latter follows from the global structure of the curvature-dependent scale function. This allows the deformation, in some parameter regimes, to affect not only the near-origin geometry but also the global spacetime profile. Within the same framework, different regions of the parameter space describe regular black holes, extremal regular black holes, and traversable wormholes.

More generally, the present results show that promoting the constant contribution entering the SV scale function to a quantity carrying information about the curvature of the vacuum counterpart provides a phenomenological route to constructing nonsingular black-bounce geometries. The specific exponential screening adopted here, motivated by the class of curvature-dependent screening functions discussed above, represents one realization of the required limiting behavior and is not intended as a unique prescription. Nevertheless, it provides a concrete setting in which the relation between vacuum curvature, short-distance saturation, the location of the bounce, and the resulting global geometry can be investigated consistently. In this sense, the construction may serve as an effective framework for exploring curvature-dependent modifications of classical black-hole geometries, while a derivation from an underlying fundamental theory and the analysis of alternative screening prescriptions remain open directions for future investigation.

An additional result concerns the role played by the topology of the transverse section. While the spherical and planar cases exhibit qualitatively similar behaviors, the hyperbolic geometry displays a considerably richer structure. In particular, the existence of regular black holes with negative mass and the strong dependence of the energy conditions on the value of the mass parameter indicate that the topology of the transverse section is not merely a mathematical detail, but rather an ingredient that significantly affects the physical properties of the matter sources and the horizon structure. The analysis of the energy conditions also provides relevant physical information. Contrary to the common expectation that the existence of a bounce necessarily requires exotic matter, we find that regular black-hole configurations may satisfy the standard energy conditions in the vicinity of the bounce for both spherical and planar topologies. This suggests that singularity avoidance can be achieved without requiring significant violations of the usual energy conditions precisely in the region where the geometry is regularized. At the same time, traversable wormhole solutions continue to exhibit the expected violations of the energy conditions near the throat, preserving one of their characteristic features. The present framework therefore provides a unified setting in which regular black holes and traversable wormholes can also be distinguished through the behavior of the matter sources supporting them.

\bibliography{mybib.bib}

\end{document}